\documentclass[aps,preprint,floatfix,nofootinbib,showpacs]{revtex4-1}
\pdfoutput=1
\usepackage{graphicx,color}
\usepackage{hyperref}

\begin{document}

{\small
\begin{flushright}
CNU-HEP-14-01
\end{flushright} }

\title{Probing the Top-Yukawa Coupling in 
Associated Higgs production with a Single Top Quark}

\def\slash#1{#1\!\!/}

\renewcommand{\thefootnote}{\arabic{footnote}}

\author{
Jung Chang$^1$, Kingman Cheung$^{1,2}$, Jae Sik Lee$^3$, and
Chih-Ting Lu$^1$}
\affiliation{
$^1$ Department of Physics, National Tsing Hua University,
Hsinchu 300, Taiwan \\
$^2$ Division of Quantum Phases and Devices, School of Physics, 
Konkuk University, Seoul 143-701, Republic of Korea \\
$^3$ Department of Physics, Chonnam National University, \\
300 Yongbong-dong, Buk-gu, Gwangju, 500-757, Republic of Korea
}
\date{\today}

\begin{abstract}
Associated production of the Higgs boson with a single top quark proceeds
through Feynman diagrams, which are either proportional to the $hWW$,
top-Yukawa, or the bottom-Yukawa couplings. It was shown in literature that
the interference between the top-Yukawa and the gauge-Higgs diagrams can
be significant, and thus the measurement of the cross sections can help
pin down the sign and the size of the top-Yukawa coupling. Here we 
perform a detailed study 
with full detector simulations 
of such a possibility at the LHC-14
within the current allowed range of $hWW$ and top-Yukawa couplings,
using $h\to b\bar b,\, \gamma\gamma,\, \tau^+\tau^-,\, ZZ^* \to 4 \ell$ modes.
We found that the LHC-14 has the potential to distinguish the size and the
sign of the top-Yukawa coupling.  
Among the channels the $h\to b\bar b$
mode provides the best chance to probe the signal, followed by 
the $h\to \gamma\gamma$ mode, which has the advantage of a narrow
reconstructed mass peak.
We also pointed out that the spatial separation among the final-state particles
has the potential in differentiating among various values of 
the top-Yukawa coupling.
\end{abstract}

\maketitle

\section{Introduction}

It has been established that the Higgs boson has been found at the Large 
Hadron Collider (LHC) \cite{atlas,cms}. The measured properties of the
Higgs boson are best described by the standard-model (SM) Higgs boson
\cite{higgcision}, which was proposed in 1960s \cite{higgs}.
The study \cite{higgcision} 
showed that the gauge-Higgs coupling $C_v \equiv
g_{hWW} = 1.01\,^{+0.13}_{-0.14}$ is very close to the SM value,
but the top- and bottom-Yukawa couplings cannot be determined as precise as 
$C_v$ by the current data. 
In particular, since the Higgs boson cannot decay into a top-quark pair, 
the top-Yukawa coupling can only be determined as $0.00 \pm 1.18$ 
($0.80 ^{+0.16}_{-0.13}$) in the fit that allows (disallows) additional 
loop contributions to $h\gamma\gamma$ and $h gg$ couplings
\footnote{ If additional
loop contributions to $h\gamma\gamma$ and $h gg$ couplings are allowed, 
the top-Yukawa coupling is only loosely bounded 
due to a very small contribution 
of associated Higgs production with a $t\bar t$ pair 
to the current Higgs data.}.
This is easy to understand because the top-Yukawa coupling only appears in the 
loops of $h\gamma\gamma$ and $hgg$ in the gluon fusion process, and also 
because the top contribution is much smaller than the $W$-loop contribution
in the $h\gamma\gamma$ coupling. Some other methods to determine the 
top-Yukawa are desired.

In literature, the most studied process of probing the top-Yukawa
is associated Higgs production with a top-quark pair 
$pp \to t \bar t h$, which can directly determine 
the absolute value of the top-Yukawa coupling.  However,
the sign cannot be determined in this process. 
On the other hand, associated Higgs production with a single top quark
has the potential of measuring the sign of the top-Yukawa coupling
\cite{tait,vernon,biswas,farina,pankaj,ellis,Englert:2014pja}.
Just take an example of 
one of the processes that contribute to a single top quark and a Higgs boson 
in the final state, $q b \to t h q^\prime$ as shown in Fig.~\ref{fig1}. 
Diagram (a) is proportional to the gauge-Higgs coupling and 
the diagram (b) is proportional to the top-Yukawa coupling.
There is another diagram with the Higgs boson attached to the bottom-quark
leg but is very small proportional to the bottom-Yukawa coupling.
The interference between the top-Yukawa diagram
and the gauge-Higgs diagram was shown to be significant and induces
large variations in the total cross section with the size and the
relative sign of the Higgs couplings to the gauge boson and the top quark. 
Therefore, if in the future the
production cross section of a single top quark and a Higgs boson can
be measured with sufficient accuracy, one can determine the size
and the relative sign of the top-Yukawa coupling.
In this work, we study associated Higgs production
with a single top quark and the potential of measuring the size and
the sign of top-Yukawa in the presence of backgrounds,
with full detector simulations.
This is the main objective of this work.

In addition to the above process, there are other processes that
a single top quark and a Higgs boson can appear in the final state:
$q g \to t h q^\prime \bar b$, $g b \to t h W^-$, 
and $q \bar q^\prime \to t h \bar b$.
Since additional or different particles appear
in the final state, all these processes can be specifically identified,
although the first process $q b \to t h q^\prime$ has the largest cross section.

In this work, we investigate various processes that contribute to the 
final states: 
$th+X$ with (i) $X=j$, (ii) $X=j+b$,
(iii) $X=W$, and (iv) $X=b$. Here top quark $t$ can decay
semileptonically or hadronically, and
the Higgs boson $h$ can decay into $b\bar b$, $\gamma\gamma$, $ZZ^* \to 4\ell$,
or $\tau^+\tau^-$. The $h\to WW^*$ mode is not considered here
because of the Higgs boson cannot be fully reconstructed.

The organization is as follows. In the next section, we lay down the 
formalism and the calculation method. In Sec. III, we show the variation
of cross sections when we vary the couplings. In Sec. IV, we calculate
the event rates with detector simulations and estimate the feasibility
at the LHC. We discuss and conclude in Sec. V.

\section{Formalism}

\begin{figure}[th!]
\centering
\includegraphics[width=5in]{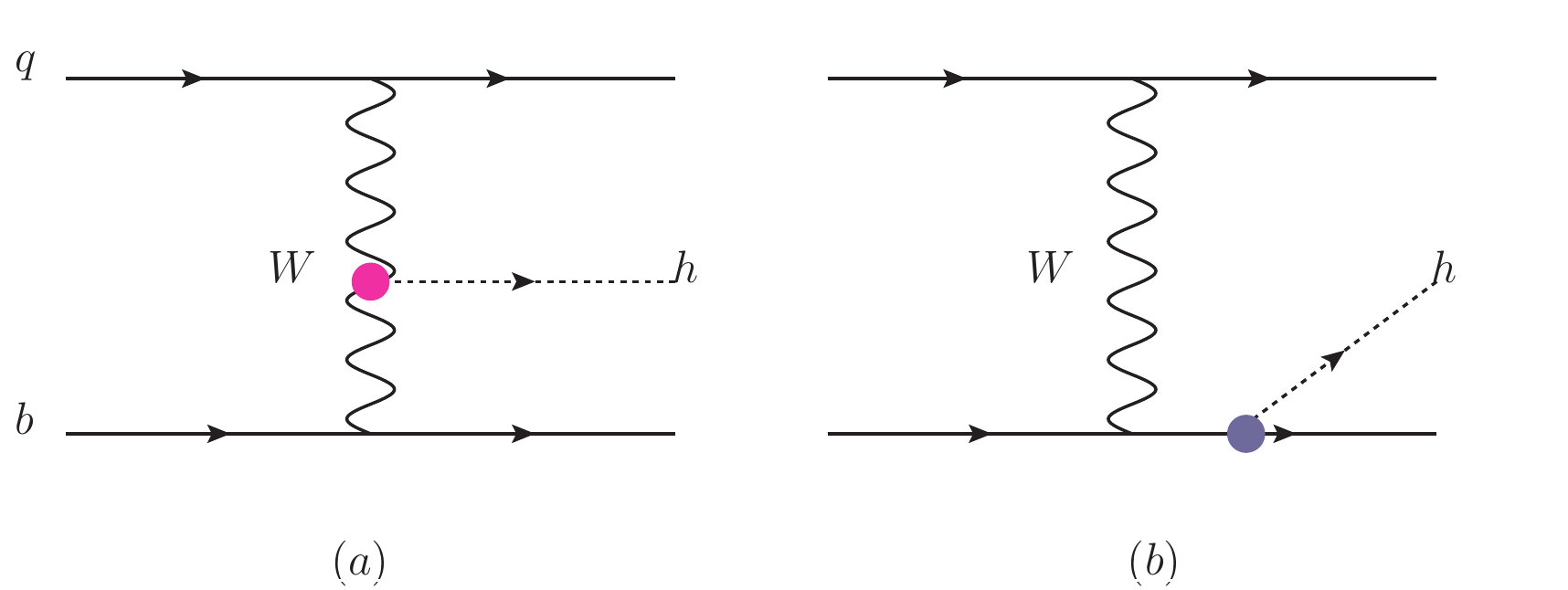}
\caption{\small \label{fig1}
Contributing Feynman diagrams for $q b \to th q^\prime$.
}\end{figure}

\begin{figure}[th!]
\centering
\includegraphics[width=5in]{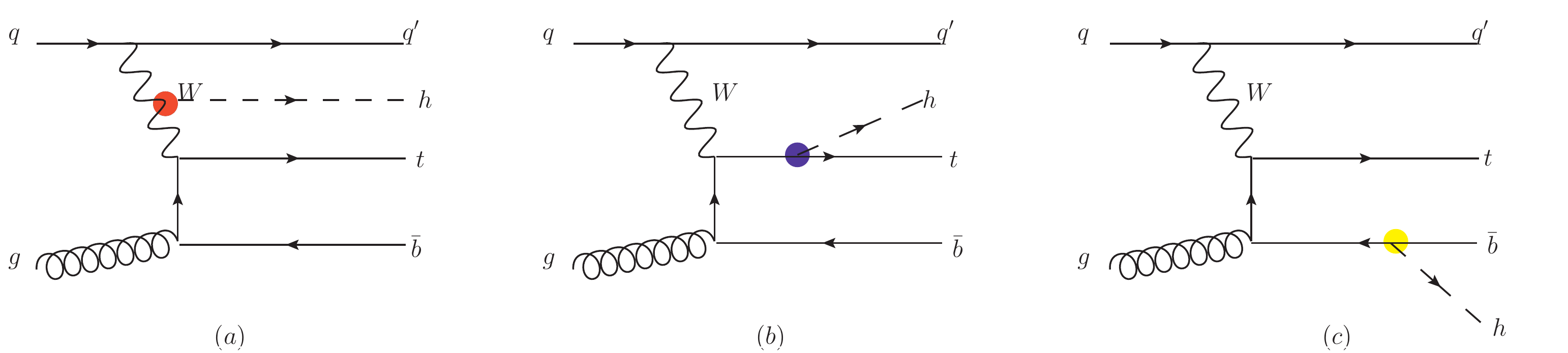}
\caption{\small \label{fig2}
Some of the contributing Feynman diagrams for $q g \to th q^\prime \bar b$.
}
\end{figure}

\begin{figure}[th!]
\centering
\includegraphics[width=5in]{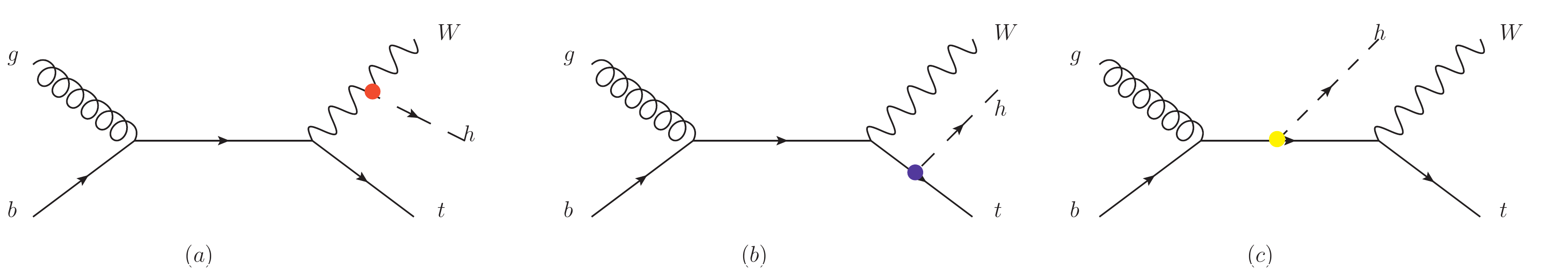}
\caption{\small \label{fig3}
Some of the contributing Feynman diagrams for $g b \to t h W^- $.}
\end{figure}

\begin{figure}[th!]
\centering
\includegraphics[width=5in]{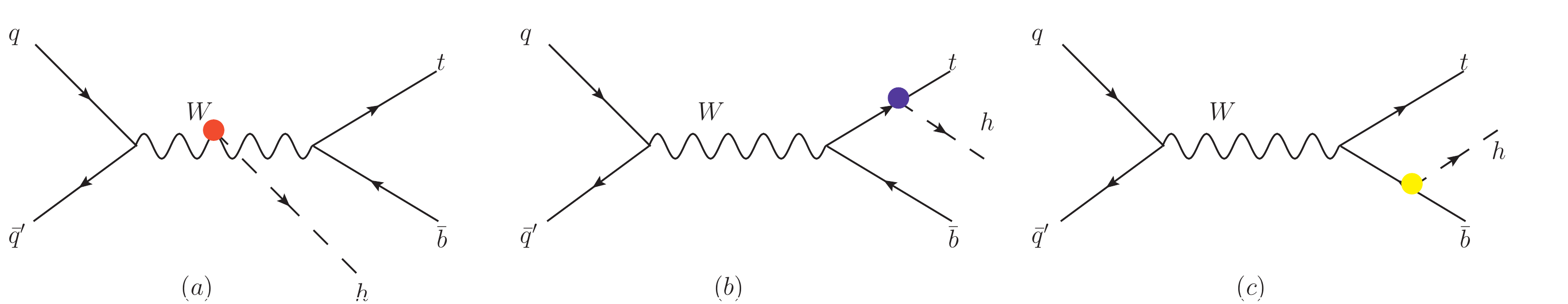}
\caption{\small \label{fig4}
Contributing Feynman diagrams for $q \bar q^\prime \to t h \bar b $.}
\end{figure}

The production processes that contribute to a single top quark and
a Higgs boson plus anything else can be found in Figs.~\ref{fig1} -- \ref{fig4}.
We have marked in particular the vertices of $hWW$, $htt$, and $hbb$. 
The production cross sections depend on 
the relative size and sign of the gauge-Higgs 
and Yukawa couplings. 
Assuming that the Higgs boson is a generic CP-mixed state, we can write 
the gauge-Higgs and Yukawa couplings as
\begin{eqnarray}
{\cal L}_{hVV} &=& g m_W \left( g_{hWW} W^+_\mu W^{-\mu} + 
 g_{hZZ} \frac{1}{2 c_W^2} Z_\mu Z^\mu \right ) h \,,  \label{hvv}\\
{\cal L}_{hff} &=& - \sum_{f = t,b,c,\tau} \, \frac{g m_f}{2 m_W} 
  \bar f \left(  g^S_{hff} + i g^P_{hff} \gamma_5 \right ) f \,h \;. \label{hff}
\end{eqnarray}
Here only $f=t,b$ are relevant to the production cross sections
of the processes in Fig.~\ref{fig1}--\ref{fig4}.  In the SM, 
$g_{hWW} = g_{hZZ} = g^S_{hff} = 1$ and $g^P_{hff}=0$.

In order to calculate the event rates we have to consider the decay
branching ratios of the Higgs boson, which depend on 
$g_{hWW}$, $g_{hZZ}$, $g^{S,P}_{htt,hbb}$,
and a few more couplings, including $h\tau\tau$, $hcc$, $h\gamma\gamma$,
and $hgg$. 
The amplitude for the decay process
$h \rightarrow \gamma\gamma$ can be written as
\begin{equation} \label{hipp}
{\cal M}_{h\gamma\gamma}=-\frac{\alpha m_{h}^2}{4\pi\,v}
\bigg\{S^\gamma(m_{h})\,
\left(\epsilon^*_{1\perp}\cdot\epsilon^*_{2\perp}\right)
 -P^\gamma(m_{h})\frac{2}{m_{h}^2}
\langle\epsilon^*_1\epsilon^*_2 k_1k_2\rangle
\bigg\}\,,
\end{equation}
where $k_{1,2}$ are the momenta of the two photons and
$\epsilon_{1,2}$ the wave vectors of the corresponding photons,
$\epsilon^\mu_{1\perp} = \epsilon^\mu_1 - 2k^\mu_1 (k_2 \cdot
\epsilon_1) / m^2_{h}$, $\epsilon^\mu_{2\perp} = \epsilon^\mu_2 -
2k^\mu_2 (k_1 \cdot \epsilon_2) / m^2_{h}$ and $\langle \epsilon_1
\epsilon_2 k_1 k_2 \rangle \equiv \epsilon_{\mu\nu\rho\sigma}\,
\epsilon_1^\mu \epsilon_2^\nu k_1^\rho k_2^\sigma$. 
Including some additional loop contributions from new particles,
the scalar and
pseudoscalar form factors, retaining only the dominant loop
contributions from the third--generation fermions and $W^\pm$,
are given by
\begin{eqnarray}
S^\gamma(m_{h})&=&2\sum_{f=b,t,\tau} N_C\,
Q_f^2\, g^{S}_{hff}\,F_{sf}(\tau_{f}) 
- g_{_{hWW}}F_1(\tau_{W}) 
+ \Delta S^\gamma \,, \nonumber \\
P^\gamma(m_{h})&=&2\sum_{f=b,t,\tau}
N_C\,Q_f^2\,g^{P}_{hff}\,F_{pf}(\tau_{f})
+ \Delta P^\gamma \,, 
\end{eqnarray}
where $\tau_{x}=m_{h}^2/4m_x^2$, $N_C=3$ for quarks and $N_C=1$ for
taus, respectively.
For the loop functions of $F_{sf,pf,1}(\tau)$, 
we refer to, for example, Ref.~\cite{Lee:2003nta}.
The additional contributions $\Delta S^\gamma$ and $\Delta P^\gamma$
are due to additional particles running in the loop. In the SM,
$P^\gamma=0$ and $g^S_{hff}= g_{hWW}=1$.
The amplitude for the decay process
$h \rightarrow gg$ can be written as
\begin{equation} \label{higg}
{\cal M}_{Hgg}=-\frac{\alpha_s\,m_{h}^2\,\delta^{ab}}{4\pi\,v}
\bigg\{S^g(m_{h})
\left(\epsilon^*_{1\perp}\cdot\epsilon^*_{2\perp}\right)
 -P^g(m_{h})\frac{2}{m_{h}^2}
\langle\epsilon^*_1\epsilon^*_2 k_1k_2\rangle
\bigg\}\,,
\end{equation}
where $a$ and $b$ ($a,b=1$ to 8) are indices of the eight $SU(3)$
generators in the adjoint representation.
Including some additional loop contributions from new particles,
the scalar and pseudoscalar form factors are given by
\begin{eqnarray}
S^g(m_{h})&=&\sum_{f=b,t}
g^{S}_{hff}\,F_{sf}(\tau_{f}) +  
\Delta S^g\,,
\nonumber \\
P^g(m_{h})&=&\sum_{f=b,t}
g^{P}_{hff}\,F_{pf}(\tau_{f}) +
\Delta P^g
\,.
\end{eqnarray}
In the SM, $P^g = 0$ and $g^S_{hff}=1$.
In the decays of the Higgs boson, we can see that the partial width
into $b\bar b$ depends on $g_{hbb}$, that into $WW^*$ and $ZZ^*$ depends
on $g_{hWW,hZZ}$, and that into $\gamma\gamma$ and $gg$ depends implicitly
on all $g_{hWW,hZZ}$, $g_{htt}^{S,P}$, $g_{hbb}^{S,P}$, and $g_{h\tau\tau}^{S,P}$.

The dependence of the production cross sections and the decay branching
ratios on $g_{hWW}$ and $g^{S,P}_{hff}$ has been explicitly shown in the above
equations. Since we are primarily interested in the relative size and sign
of the gauge-Higgs and top- and bottom-Yukawa couplings, 
for bookkeeping purposes we use the following notation
\begin{equation}
\label{eq:notation}
C_v  \equiv g_{hVV}=g_{hWW}=g_{hZZ}\,, \qquad
C^{S,P}_t  \equiv g^{S,P}_{htt}\,, \qquad
C^{S,P}_b  \equiv g^{S,P}_{hbb}\,.
\end{equation}
We will show the variation of the cross sections in the next section.

\section{Variation of Cross Sections}

In this section, we show the cross sections of the processes listed
in the last section versus the top-Yukawa $C_t^S$ and $C_t^P$. 
We use MADGRAPH \cite{madgraph} with the 5-flavor scheme ($u,d,s,c,b$ partons) 
for calculating the cross sections. We do not impose cuts as we are
presenting the total cross sections here, except for the 
process $pp \to th j b$, where we have to impose cuts on the 
final state $b,j$ to remove the divergences.
We use CTEQ6 \cite{cteq} for parton distribution functions with the
renormalization/factorization scale equal to $M_Z$.

\begin{figure}[th!]
\includegraphics[width=3.2in]{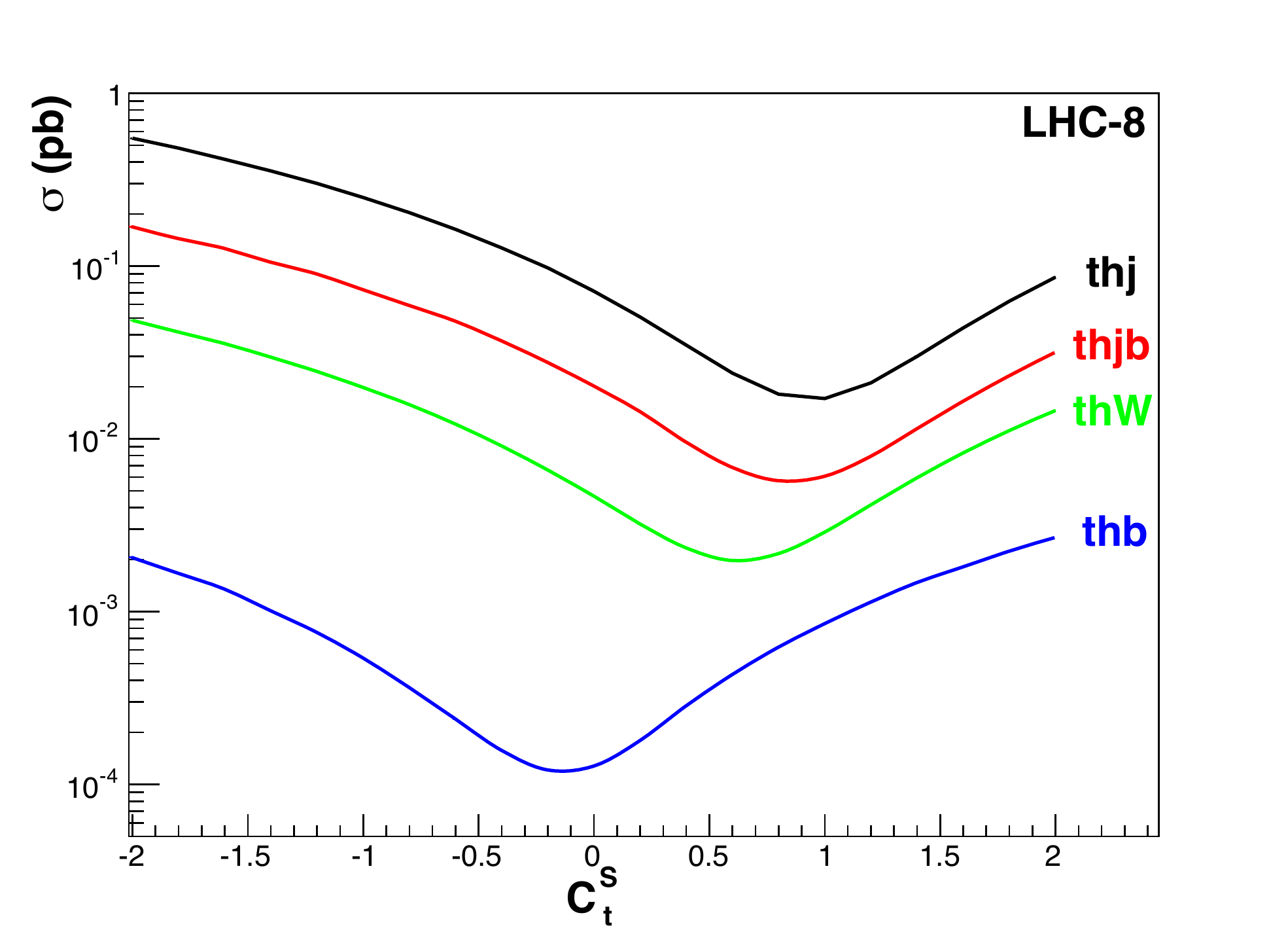}
\includegraphics[width=3.2in]{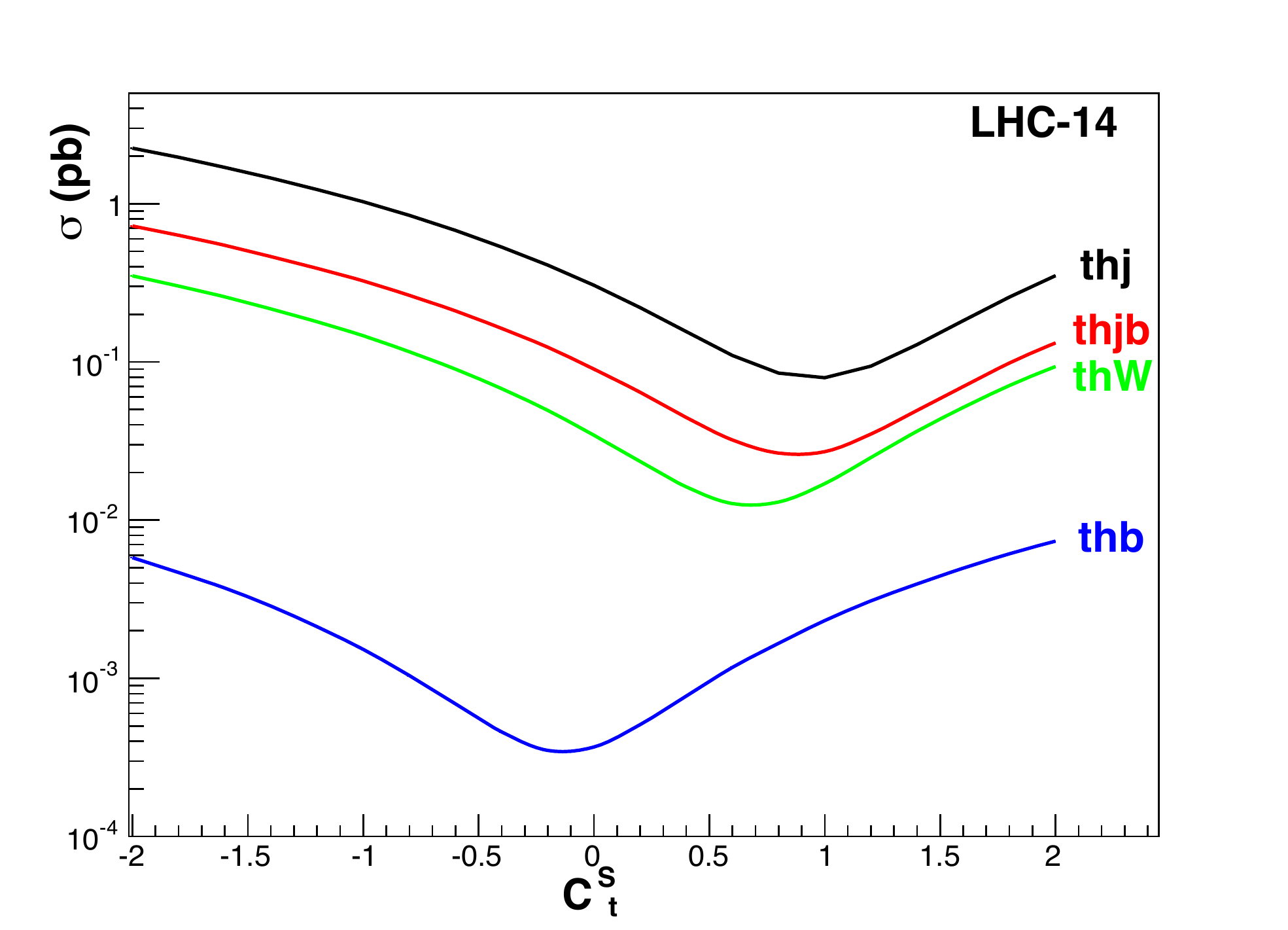}
\caption{\small\label{prodcx} 
Variation of the total cross sections versus $C_t^S$ for $pp \to t h X $ 
with $X=j, jb, W, b$ in the order of the size of cross sections
at (a) LHC-8 and (b) LHC-14.
We have taken $C_v=C_b^S=1$ and $C_{t,b}^P=0$. No cuts are imposed
except for the second process $pp \to th jb$ in which we applied
the cuts in Eq.~(\ref{bcut}) to remove the divergence.
}
\end{figure}

\subsection{$q b\to t h q^\prime$}
The Feynman diagrams are shown in Fig.~\ref{fig1}. We have also included
the subprocesses with $\bar b$ and all possible $q$ and $\bar q$ 
in the initial state. Therefore, both $th j$ and $\bar t hj$ final states
are included. It is clear that 
both couplings $C_v$ and $C_t^{S,P}$ can affect the total cross sections, 
while $C_b^S$ has a negligible effect because of the small $b$-quark mass.

In Fig.~\ref{prodcx} we show the total cross sections 
of $pp \to th j$ (the curve at the top) versus $C_t^S$ 
at the LHC-8 and LHC-14, and we have fixed $C_v=1$ and $C_t^P=0$.
It is clear that the cross section depends crucially on the value
of $C_t^S$. The minimum cross section of $thj$
appears very near the SM value of $C_t^S=1$.
The cross section keeps increasing for $C_t^S$ decreasing from 1 and for
$C_t^S$ increasing from 1. 
On the other hand, the effect of $C_b^{S,P}$ 
on the production cross section is very small
\footnote{
The cross section multiplied the branching ratio $\sigma (pp\to t h j)\times
B(h \to b\bar b)$ strongly depends on $C_b^{S,P}$, because the partial
width into $b\bar b$ is directly proportional to 
$\beta^2|C_b^S|^2+|C_b^P|^2$ with $\beta^2=1-4m_b^2/m_h^2$.
Other processes have similar features with
the variation in $C_b^{S,P}$. The production cross section itself shows very small
effects from $C_b^{S,P}$, but the cross section multiplied by the branching
ratio $\sigma \times B(h \to b\bar b)$ varies significantly with $C_b^{S,P}$.
}.
In addition to the figure, we also show the cross sections for these
four processes in Table~\ref{procx-table} for $C_t^S = 1,0,-1$.
It is clear that the size of the cross sections decreases as
$ X=j > X=jb > X=W > X =b$.

\subsection{$g q \to t h q^\prime \bar b$}
Some of the contributing  Feynman diagrams are shown in Fig.~\ref{fig2}. 
Both $t h j \bar b$ and $\bar t h j b $ final states
are included. 
Note that one can regard this process as a higher-order correction
to the process 
$q b\to t h q^\prime$ in the previous subsection
when we do not tag the $b$-quark in the final state.
In order to distinguish them, we impose a minimal set of cuts on the
$b$ and $j$ (also needed to avoid the collinear divergence):
\begin{equation}
\label{bcut}
  p_{T_b} > 25 \; {\rm GeV}\,, \qquad |\eta_b| < 2.5\,; \qquad
  p_{T_j} > 10 \; {\rm GeV}\,, \qquad |\eta_j| < 5\, .
\end{equation}
Also, this set of cuts is used to avoid the double-counting of the
cross section against the process
$q b\to t h q^\prime$ in the previous subsection.
We show the variation of the cross sections 
of $pp \to thjb$ (the second curve from the top) versus $C_t^S$ at the LHC-8
and LHC-14 in Fig.~\ref{prodcx}. The minimum cross section occurs at 
about $C_t^S = +0.85 $, and the cross section increases approximately symmetric
about this minimum point.

\subsection{$g b\to t  h W $}
Some of the contributing Feynman diagrams are shown in Fig.~\ref{fig3}. 
Both $t h W^-$ and $\bar t h W^+$ final states are included. 
We show the variation of the cross sections of 
$pp \to thW$ (the third curve from the top) versus  $C_t^S$ at the LHC-8
and LHC-14 in Fig.~\ref{prodcx}. The minimum cross section occurs at 
about $C_t^S = +0.6 $, and the cross section increases approximately symmetric
about this minimum point.

\subsection{$q \bar q^\prime \to t h \bar b $}
The Feynman diagrams are shown in Fig.~\ref{fig4}. 
Both $t h \bar b$ and $\bar t h b$ final states
are included. 
We show the variation of the cross sections of
$pp \to thb$ (the bottom curve) versus $C_t^S$ at the LHC-8
and LHC-14 in Fig.~\ref{prodcx}. The minimum cross section occurs at 
about $C_t^S = -0.15$,  which is far from the SM value, 
and the cross section increases approximately symmetric
about this minimum point.
Being different from the three processes considered before,
$q \bar q^\prime \to t h b$ is an $s$-channel process mediated
by a mostly off-shell $W$.

\begin{table}[t!]
\caption{\small \label{procx-table}
The leading-order production cross sections in fb for the processes
$pp\to t h+X$ at 14 TeV (8 TeV) LHC, taking $C_v=C_b^S=1$ and $C_{t,b}^P=0$. 
We have not applied any cuts 
except for the case with $X=j+b$ for which we required
$p_{T_b} > 25 \; {\rm GeV}\,, |\eta_b| < 2.5\,; 
p_{T_j} > 10 \; {\rm GeV}\,, |\eta_j| < 5$, see text for details.  }
\medskip
\begin{ruledtabular}
\begin{tabular}{lcccc}
& \multicolumn{4}{c}{$\sigma(pp\to thX)$[fb]} \\
   & $X=j$ & $X=j+b$ & $X=W$ & $X=b$ \\
\hline
$C_t^S=+1$ (SM) & 79.4 (17.1) & 27.1 (5.95) & 17.0 (2.89) &  2.32(0.833)  \\
$C_t^S=0$ & 305 (71.4) &  90.0 (19.8) & 34.4 (4.66) &  0.368 (0.126)  \\
$C_t^S=-1$ & 1030 (249) &  325 (72.8) & 146 (19.8) & 1.52 (0.536) \\
\end{tabular}
\end{ruledtabular}
\end{table}

\subsection{Variation of the cross sections versus $C_t^P$}

\begin{figure}[th!]
\centering
\includegraphics[width=3.2in,height=3in]{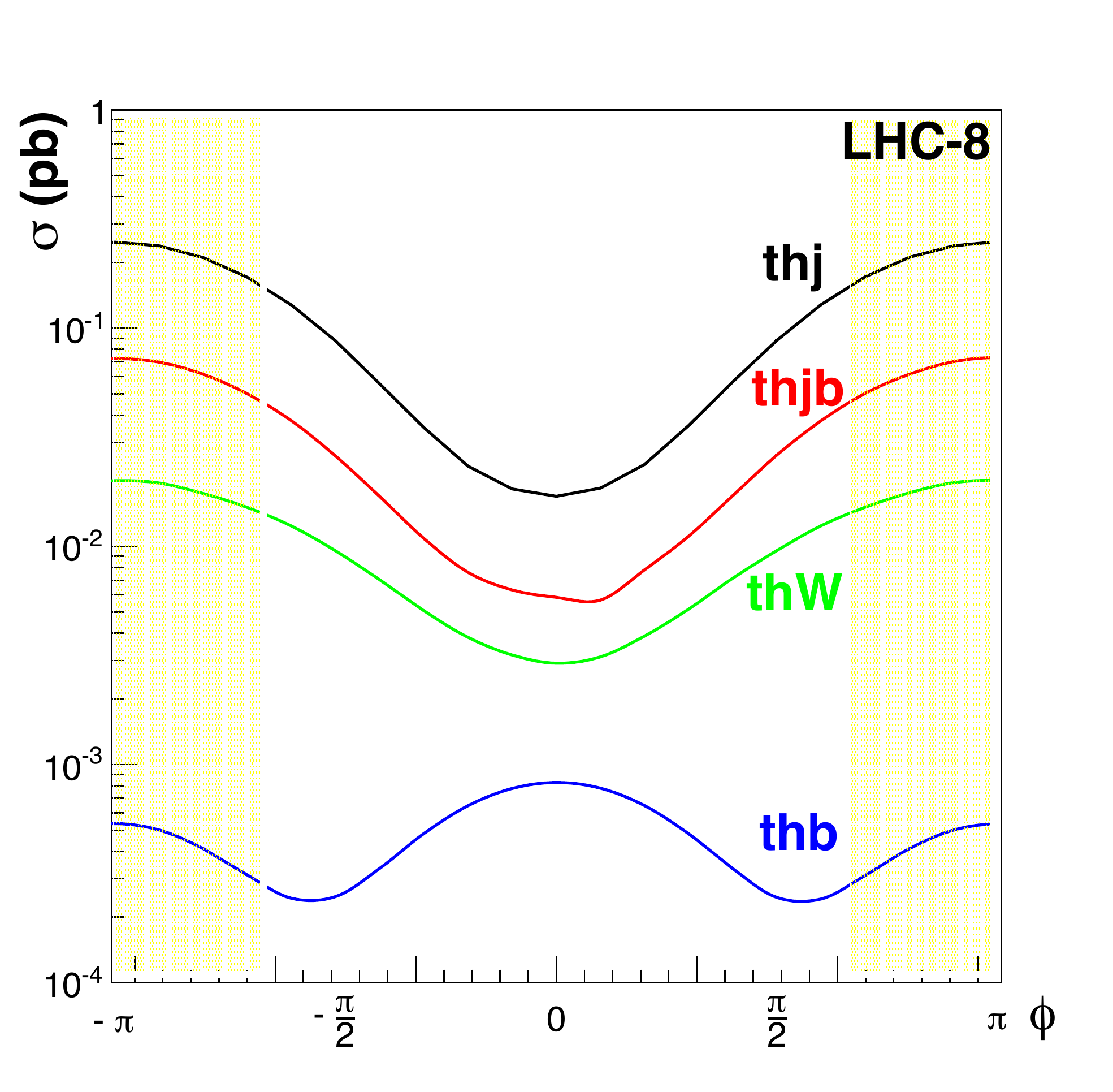}
\includegraphics[width=3.2in,height=3in]{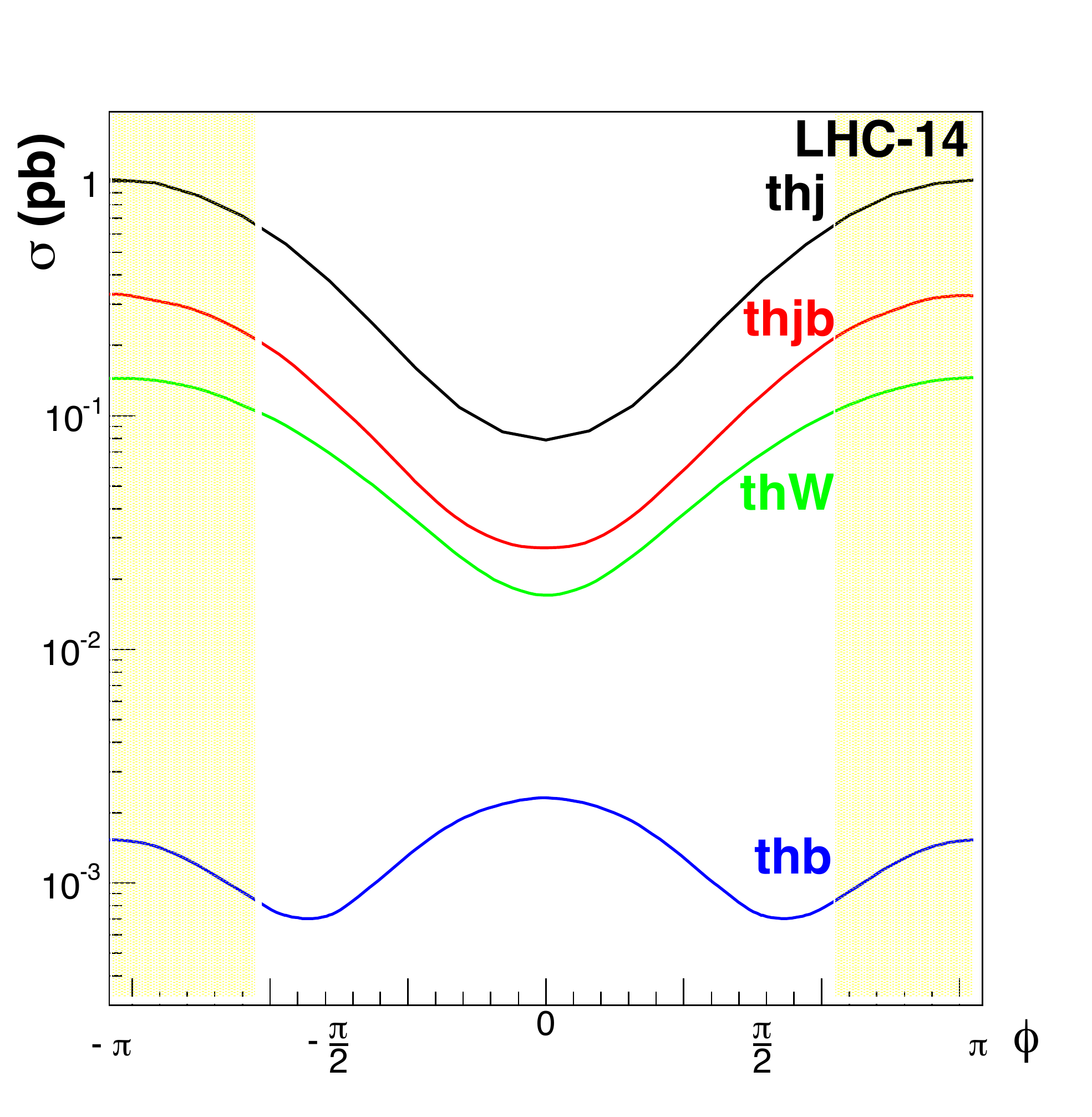}
\caption{\small \label{prod-cp}
Production cross sections at the LHC-14 
for $pp \to t h j$ versus $\phi=\arctan(C_t^P/C_t^S)$
under the constraint 
$\left( C_t^S/0.86 \right)^2+\left( C_t^P/0.56 \right)^2=1$.
We take $C_v=1$. 
The shaded regions are those disallowed at 68\% C.L. by the
Higgs data obtained in Ref.~\cite{higgcision}.
}
\end{figure}

So far we only concern the scalar component in the top-Yukawa
coupling. In Eq.~(\ref{hff}), we can also have the
pseudoscalar component in the coupling, 
which is proportional to $i \gamma^5 g_{hff}^P$. 
In this subsection, we examine the variation of the cross sections when
the pseudoscalar component is present in the $h t\bar t$ vertex.
It was shown in Ref.~\cite{higgcision} that the scalar and pseudoscalar
components in $ht\bar t$ are constrained nontrivially, as shown in 
Fig. 10(c) of Ref.~\cite{higgcision}. The $C_t^S$ and $C_t^P$ are roughly
constrained by an elliptical equation, given by \cite{higgcision}
\[
   1 = \frac{\left(C_t^S \right)^2}{ (0.86)^2 } 
     + \frac{ \left( C_t^P \right)^2} {(0.56)^2 } \;.
\]
We can parameterize $C_t^S$ and $C_t^P$ by
\begin{equation}
 C_t^S = 0.86 \, \cos\theta  \,; \ \ \
 C_t^P = 0.56 \, \sin\theta \;.
\end{equation}
The actual angle $\phi$ presented in the plane of
$(C_t^S, C_t^P)$ is related to $\theta$ by 
\[
 \tan \phi \equiv \frac{C_t^P}{C_t^S} =
    \frac{0.56 \sin\theta}{0.86 \cos\theta} = 0.66 \, \tan\theta \;,
\]
where the ranges of $\phi$ is $- \pi \le \phi < \pi$. 
Nevertheless, if we restrict to the 68\% C.L. region of the Fig.10(c) of 
Ref.~\cite{higgcision}, the range of allowed $\phi$ 
is approximately $-2\pi/3 \le \phi \le 2\pi/3$.
We show the cross sections versus $\phi$ in Fig.~\ref{prod-cp},
in which the shaded regions are those disallowed at 68\% C.L. obtained
in Ref.~\cite{higgcision}.
It is interesting to note that the first three curves at the top of the
figure have similar behavior across $\phi$ while the bottom curve has
the opposite behavior. Again it is due to the $s$-channel exchange mediated
by a mostly off-shell $W$ in the last process.

One comment about the next-to-leading order (NLO) corrections is in order here.
Since the NLO QCD corrections to single-top
plus Higgs production are very similar to single-top production, we can
roughly estimate the QCD corrections to the current processes by
looking up the NLO corrections to single-top production.  A
number of NLO and next-next-to-leading order calculations existed in
literature for single top-quark production \cite{nlo}.  The NLO
corrections to the process $q b \to t q' $ and $qg \to t q' \bar b$ are
very modest, usually less than 10\%, while those of $g b \to t W^-$ and
$q\bar q' \to t \bar b$ can be as large as $40-50\%$.
We shall estimate the potential at the LHC using the process 
$q b\to t hq'$, which has the largest cross section among the signal
processes, and therefore the NLO correction on the signal cross section
is a mere less than 10\% effect.

\section{Potential at the LHC}
We have demonstrated in the previous section that when we change
$C_t^S$ the production cross sections change significantly. If one can
measure the event rates of associated Higgs production with a single top
quark, the size and sign of $C_t^S$ can be determined.  
There are 4 production processes of a Higgs boson and a single top
at the LHC: $pp\to thX$ with $X=j$, $X=j+b$,  $X=W$, and $X=b$.
The top quark and the Higgs boson decay subsequently. Including the semileptonic
and hadronic decays of top quark and the five Higgs decay modes into
$b\bar b$, $\tau^+\tau^-$, $W W^*$, $Z Z^*$, and $\gamma \gamma$,
we have 40 channels which require different search strategies against 
different backgrounds.

In this section, we
look at a few decay channels of the Higgs boson and investigate the 
feasibility of isolating the signal events in the presence of backgrounds
after implementing detector simulations.
A few decay channels that we shall study are: 
$h\to b\bar b$, $h \to \gamma \gamma$, $h \to\tau^+\tau^-$, and 
$h\to Z Z^* \to 4 \ell$.
These are the channels enable one to reconstruct the
Higgs boson, especially the 
$h \to \gamma \gamma$ and $h\to Z Z^* \to 4 \ell$, which can help reducing the
backgrounds by imposing the invariant-mass cut on $M_{\gamma\gamma}$ or
$M_{4\ell}$. The other two channels $h\to \tau\tau, b\bar b$ are not as 
effective as $\gamma\gamma$ and $ZZ^*\to 4\ell$ channels in reconstructing
the invariant mass.  We delay the channel $h \to W W^* \to \ell \nu \ell \nu$
to later studies.
For the top quark decay we can choose either the semileptonic or hadronic decay,
depending on how complicated the final state will be.  For example,
if $h\to b\bar b$ we only choose the semileptonic decay for the top quark.
If we choose a non-hadronic decay mode for the Higgs boson, we can afford
the luxury to have both the semileptonic and hadronic decays of the top quark.

We calculate the signal and background processes and 
generate events by MADGRAPH \cite{madgraph},
perform parton showering by Pythia \cite{pythia}, 
and employ the detector simulations by Delphes 3 \cite{delphes3}.
We will give details about the selection cuts, detection efficiencies, and 
signal and background event rates in the next few subsections.
For easy reading we summarize the detection efficiencies for $b$ quarks 
\cite{cms-b}, $\tau$ leptons \cite{cms-tau1,cms-tau2}, 
charged leptons ($\mu$ and $e$), and photons in 
Table~\ref{eff-table}, as well as the mis-tag probabilities for 
the charm quark to fake a $b$-jet, other light quarks or gluon
to fake a $b$-jet \cite{cms-b}, mis-tag probability for a jet to fake a
tau lepton, and that of a jet to fake a photon \cite{jet}
\footnote
{The $B$ tagging efficiency can be as high as 85\% but at that point
the corresponding mis-tag probability goes up quickly \cite{cms-b}.
We take moderate values for both quantities in this study. 
If we take a smaller $\epsilon_b = 0.6$, the mis-tag probabilities goes
down as $P_{udsg \to b} = 0.004$ and $P_{c \to b} = 0.08$.  }.
For simplicity we assume the efficiencies
are constant over a large range of transverse momentum $p_T$ larger than
the acceptance cut (e.g. $p_T > 25$ GeV). The efficiencies for charged
leptons ($e$ and $\mu$) and photons are more than 90\%, and so we 
simply assume them to be $1$.

We mainly focus on the production with the subprocess
$q b \to t h q^\prime$ (Fig.~\ref{fig1})
because its cross section is the largest. The processes
in Figs.~\ref{fig3} and \ref{fig4} are much smaller at the LHC-14.
The process in Fig.~\ref{fig2} is similar to the first process, and has
a cross section about 20-30\% of that of the first process and also
one additional $b$ quark in the final state. 
Specifically, we consider the processes:
$pp\longrightarrow t(\to bl\nu)+h(\to bb\,,\gamma\gamma\,,\tau^+\tau^-)+j$ 
and
$pp\longrightarrow t(\to b j_1j_2)+h(\to  Z Z^* \to 4 \ell)+j$.
We find that the hadronically decaying top channel with 
$h\to\gamma\gamma$ is less efficient than the semileptonic one and
we present only the latter case.

\begin{table}[tbh!]
\caption{\small \label{eff-table}
The detection efficiencies taken in this work 
for $b$ quarks, $\tau$ leptons, charged leptons 
($\mu$ and $e$), and photons, as well as the mis-tag probabilities for other
light quarks to fake a $b$-jet 
or a $\tau$ lepton \cite{cms-b,cms-tau1,cms-tau2}.
We also list the probability for a jet faking a photon \cite{jet}.
The $B$ Tagging efficiency and mis-tag probabilities are correlated.
The numbers in parenthesis are for $\epsilon_b = 0.6$. }
\medskip
\begin{ruledtabular}
\begin{tabular}{cccc}
\multicolumn{4}{c}{Detection efficiencies}\\
\hline
$\epsilon_b$ & $\epsilon_\tau$ & $\epsilon_\ell$ & $\epsilon_\gamma$ \\
$0.7 \; (0.6) $        & $0.5$        & $1.0$        & $1.0$        \\
\hline
\multicolumn{4}{c}{Mistag probability}\\
\hline
$P_{c\to b}$ & $P_{udsg \to b}$ &  $P_{j \to \tau}$ &  $P_{j \to \gamma }$  \\
$0.2\; (0.08)$       & $0.015 \; (0.004)$  & $0.01$     & $10^{-3}$
\end{tabular}
\end{ruledtabular}
\end{table}

\subsection{Semileptonic top decay}
In this subsection, we consider top-Higgs associated production $thj$ 
with the single top decaying semileptonically
\begin{equation}
pp \longrightarrow \, thj \to (bl\nu)\,h\,j \;.
\end{equation}
At this stage, we apply a set of  basic cuts
\begin{eqnarray}
\label{eq:semitop-cut1}
&& \Delta R_{ij} > 0.4  \qquad \mbox{with $i,j$ denoting $b$, $j$, and $\ell$}\,,
\nonumber \\
&& p_{T_b} > 25 \,{\rm GeV}\,, \;\; |\eta_b | < 2.5 \,,
\nonumber \\
&&   p_{T_\ell} > 25 \,{\rm GeV}\,, \;\;  |\eta_\ell | < 2.5 \,,
\nonumber \\
&& p_{T_j} > 25 \,{\rm GeV}\,,\;\;  |\eta_j | < 4.7\,.
\end{eqnarray}
The spatial separation among the objects (the $b$ jets, the jet, and 
the lepton) in the final state is denoted by $\Delta R_{ij}$.

\begin{figure}[th!]
\centering
\includegraphics[width=3.2in]{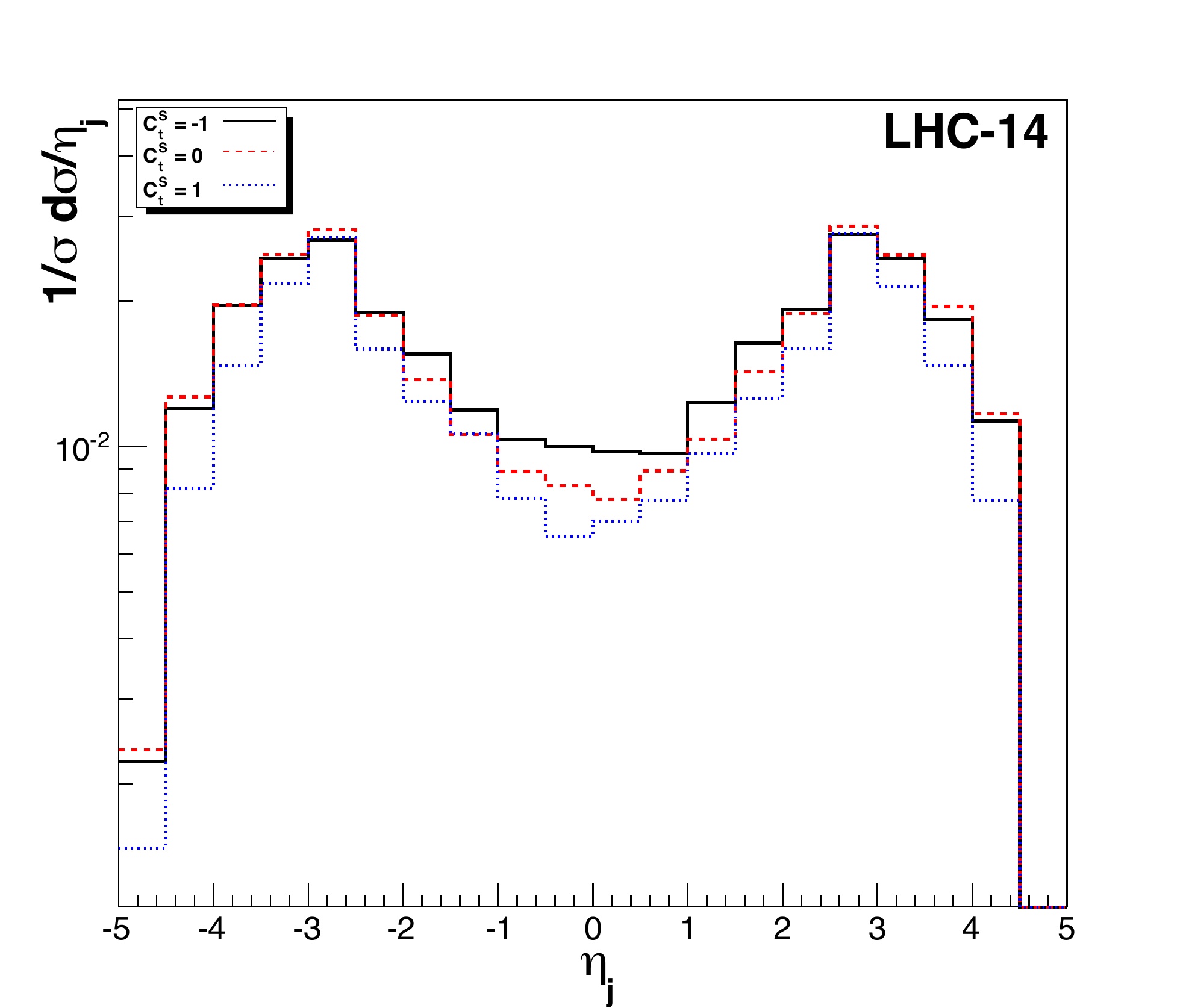}
\includegraphics[width=3.2in]{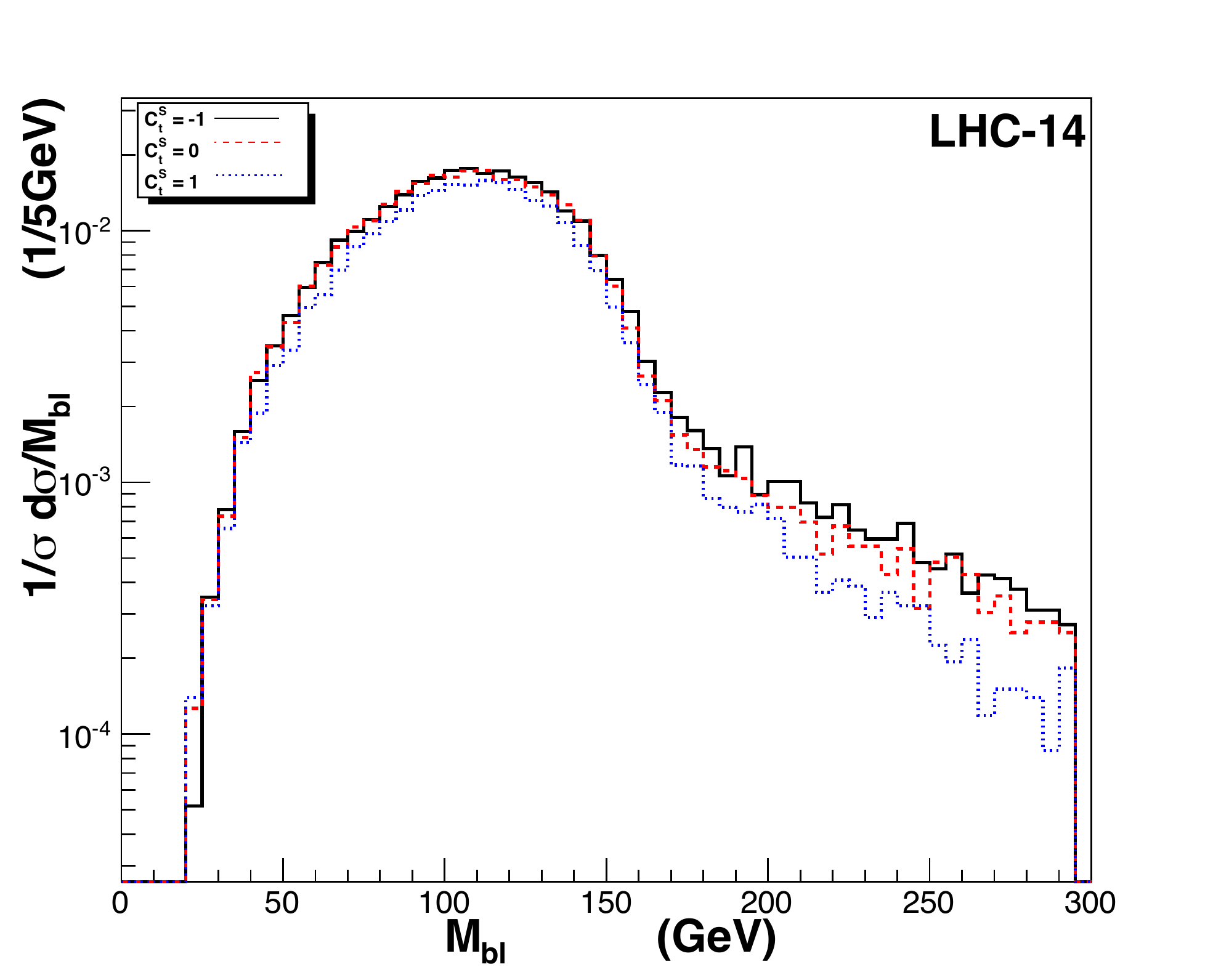}
\caption{\small \label{dist}
(a) Normalized spectra for the $\eta_j$ distribution and (b) 
normalized invariant mass spectra $M_{bl}$ of the
process $pp \to th j$ at the LHC-14. We have imposed the set of basic cuts
in Eq.~(\ref{eq:semitop-cut1}) with detector simulations.
}
\end{figure}

If we look at the Feynman diagrams of the subprocess $qb \to t h q'$ 
(Fig.~\ref{fig1}), the dominant contribution comes from where 
the intermediate $W$ is almost on-shell, which implies that the
incoming $q$ behaves like a spectator and therefore it tends to go
forward. This behavior is similar to those encountered in $WW$ scattering
\cite{wwscatter}. In Fig.~\ref{dist}(a), we show the spectra
of the pseudorapidity of the forward jet for $C_t^S = 1,0,-1$. All the
three curves indeed show the forward behavior.  We therefore impose
the forward-jet requirement.
Another useful cut is on the invariant mass of the $b$ quark and the
charged lepton $\ell$ coming from the top quark decay. Thus, the
invariant mass should always be less than $m_t$. We show in 
Fig.~\ref{dist}(b) the invariant mass spectra $M_{bl}$ for $C_t^S = 1,0,-1$
with detector simulation.
After the set of basic cuts listed in Eq.~(\ref{eq:semitop-cut1}),
we further require the forward jet-tag and invariant mass cut on $M_{bl}$,
given by
\begin{eqnarray}
\label{eq:semitop-cut2}
2.5 < |\eta_j | < 4.7,  \nonumber \\
M_{bl} < 200 \, {\rm GeV} \;.
\end{eqnarray}
The next level of cuts will depend on the decay channel of the Higgs boson.

\subsubsection{$h \to b \bar b$}
We start with the most difficult decay mode of the Higgs boson because of 
the large hadronic background. 
%
%
Therefore, with the first process in mind the final state consists
of 3 $b$ quarks, one charged lepton, and a missing energy due to the neutrino
\footnote{
In Ref.~\cite{farina}, the process $pp \to th jb$ was studied with 
the final state containing $4b$ jets.
 They showed that the sensitivity 
is better than that of the process $pp \to th j$ with $3b$ jets
 in the 
final state. However, a full detector simulation is needed to establish
the statement.  }.
The charged lepton with the corresponding $b$ quark from the top decay
can be selected with high purity by choosing the smallest $M_{b \ell}$
among the three combinations.
The other two $b$ quarks are then considered the $b$ quarks coming 
from the Higgs boson decay, and can be reconstructed at the Higgs boson mass.
For all three $b$ quarks including those from decaying Higgs,
we impose the same cuts on their momenta and rapidities
as in Eq.~(\ref{eq:semitop-cut1}).

The major reducible and irreducible backgrounds are QCD production of
\begin{eqnarray}
&& (i)\ t\bar t \to  t (\bar b j_1 j_2) \to t b \bar b j  \ \
\mbox{with mis-tagging one of $j_1$ and $j_2$ as $b$, } 
\nonumber \\
&& (ii)\ t\bar t j \to  t (\bar b j_1 j_2) j \to t b \bar b j \ \
\mbox{with mis-tagging one of $j_1$ and $j_2$ as $b$ and missing the other one,} 
\nonumber \\
&& (iii)\ t b \bar b j\,, \ \ \mbox{and} 
 \ \ (iv)\ tZj \to  t b \bar b j \nonumber
\end{eqnarray}
In addition to the basic cuts as in 
Eqs.~(\ref{eq:semitop-cut1}) and (\ref{eq:semitop-cut2}),
we further impose the following selection cuts:
\begin{eqnarray}
\label{eq:semitop-cut3}
&& |M_{b_1\bar b_2} - m_h | < 15 \,{\rm GeV}\,, \nonumber  \\
&& M_{b_1\bar b_2 j} > 300\, {\rm GeV}\,,
\end{eqnarray}
to separate the signal events from backgrounds.
Here,
$b_{1,2}$ denote the bottom quarks which are supposedly coming from 
the Higgs boson in the signal process while we
identify the bottom quark $b$ from the decaying top
with the smallest $M_{bl}$, as we have mentioned above, 
on which we then put the cut $M_{bl} < 200$ GeV.

We require the correct $b_1\bar b_2$ pair to satisfy
the Higgs mass window of $\pm 15$ GeV.  We note that we cannot take a 
smaller window because of the wide spreading of the Higgs peak with
detector simulation, in contrast to parton-level studies. We will show 
the invariant mass spectrum of the $b\bar b$ pair shortly.
The forward jet-tag is used because of the forward nature of the accompanying
jet in the signal process. Finally, we used a cut on the invariant mass
$M_{b_1 \bar b_2 j}> 300$ GeV of 
the $b\bar b$ pair coming from the Higgs decay and the accompanying jet.

\begin{table}[tb!]
\caption{\label{hbb} \small
The cut flow of cross sections in fb  
at the LHC-14 for the signal $pp \to t h j$ 
with semileptonic decay of the top quark and $h\to b \bar b$,
and various  backgrounds.  We have used
the $B$-tag efficiency $\epsilon_b = 0.6$, mis-tag $P_{c \to b} = 0.08$
and $P_{udsg \to b} = 0.004$.
}
\medskip
\begin{ruledtabular}
\begin{tabular}{lccccc}
Cuts & \multicolumn{3}{c}{Signals (fb)} & \multicolumn{2}{c}{Backgrounds (fb)}\\
     & $C_t^S=1$ & $C_t^S=0$ & $C_t^S=-1$ & $t\bar t$ & $t \bar t j$ \\
\hline
(1) Basic cuts Eq.~(\ref{eq:semitop-cut1}) and & & & & \\
{} $p_{T_{b_{1,2}}} > 25$ GeV, $|\eta_{b_{1,2}}|<2.5$ 
 &  0.793    & 4.23 & 15.29  &   655  &  797  \\
(2) $2.5 < |\eta_j|<4.7$ &   0.388   & 2.20 & 7.68 &  46.2  & 95.6 \\
(3) $(M_{bl})^{\rm min} < 200$ GeV & 0.387& 2.19 & 7.59 & 46.2 & 95.6  \\
(4) $|M_{b_1 b_2} - m_h | < 15$ GeV  & 
   0.13 & 0.74 & 2.5 &  6.69 & 15.2   \\
(5) $M_{b_1 b_2 j}>300$ GeV &  0.06 & 0.3  &  0.9 &  1.34  &   5.41  \\
\hline
\hline
$S/\sqrt{S+B}$ for 300 fb$^{-1}$ & 0.40  &  2.0 &  5.6
\end{tabular}
\end{ruledtabular}
\end{table}

\begin{figure}[tb!]
\centering
\includegraphics[width=5in]{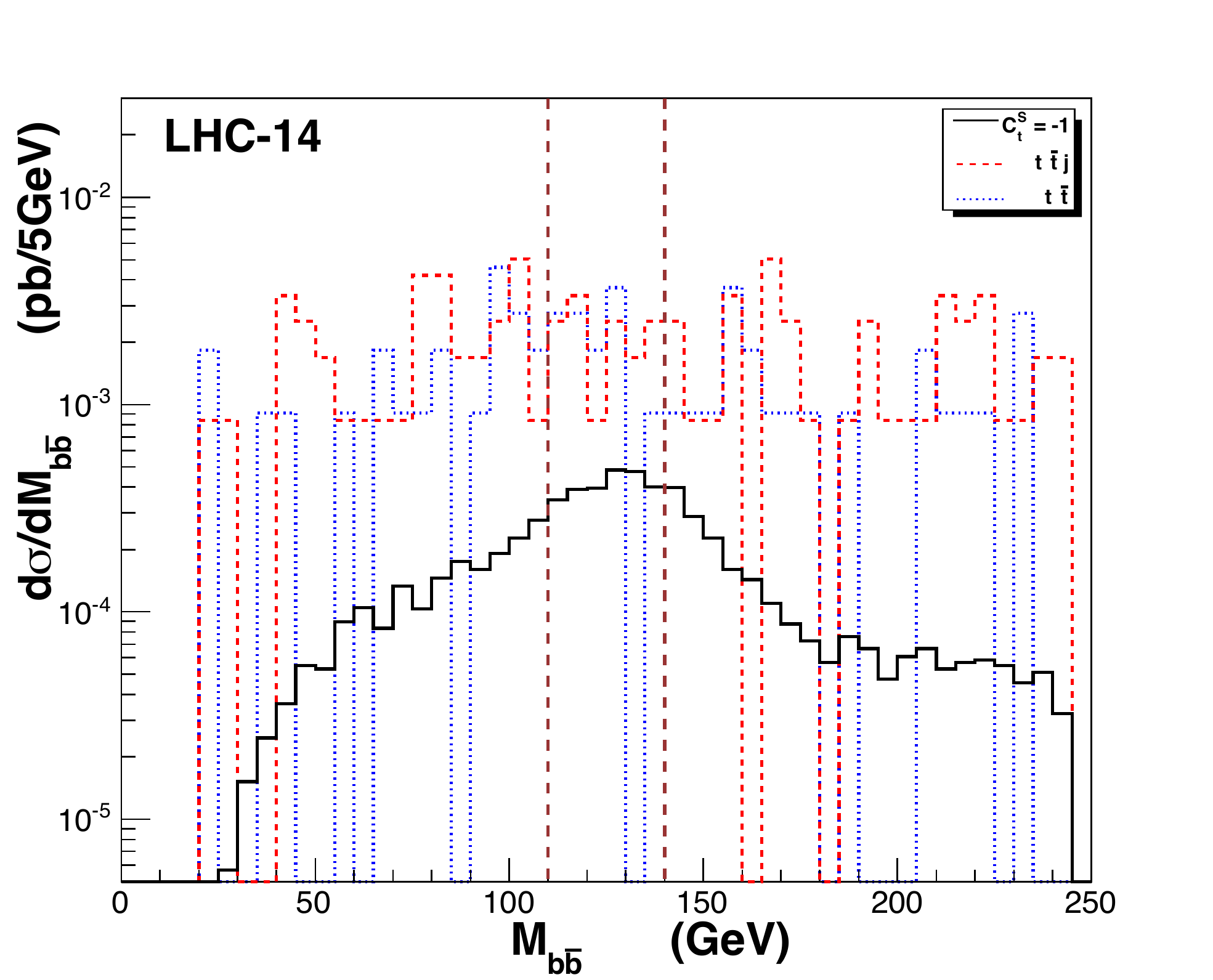}
\caption{\small \label{fig-bb}
Invariant mass $M_{b\bar b}$ distributions for the signal
$pp \to t h j$ with $C_t^S = -1$ followed by the semileptonic decay
of the top quark and $h\to b \bar b$, 
and for the $t\bar t$ and $t\bar t j$ backgrounds
at the LHC-14. The shape of the peak for $C_t^S=1,0$ is the same as the
one shown with $C_t^S = -1$.  The selection cuts up to level (3) of 
Table~\ref{hbb} have been applied.
The two vertical dashed lines are $m_h \pm 15$ GeV.
 }
\end{figure}

We show the cut flow of cross sections for the
signals and backgrounds at the LHC-14 in Table~\ref{hbb}.
The cross sections shown are calculated with the $B$-tagging efficiency 
$\epsilon_b = 0.6$ and mis-tag probabilities $P_{c\to b} =0.08$ and
$P_{udsg \to b} = 0.004$. We found that the set of probabilities with
$\epsilon_b=0.7$ would give a somewhat smaller significance, because of
much larger mis-tag probabilities.
We only include
the two most significant backgrounds in this study, namely, the $t\bar t$
and $t\bar t j$ backgrounds. The other few backgrounds ($tb\bar b j$, 
$tZj$, etc) are substantially smaller than these two and so
would not affect the estimates of significance here.  The $t\bar tj$ 
background turns out to be the largest background in this channel, because
the addition jet in the matrix-element level can be highly energetic,
in contrast to the jet activities coming from showering.

One of the most crucial cuts is the invariant mass cut on the $b\bar b$
pair coming from the Higgs boson decay. In parton-level, this cut would 
be 100\% efficient for the signal and can cut away a very large fraction of the
backgrounds. However, with detector reconstruction the invariant mass peak is
largely spread out so that we cannot employ a very narrow cut. 
The invariant mass distributions are shown for the signals and backgrounds
in Fig.~\ref{fig-bb}. Another interesting cut is the forward jet cut, as
have been explained above. Finally, the cut on $M_{b\bar bj}$ with the 
$b\bar b$ pair from Higgs-boson decay is also effective in reducing the
backgrounds. 

In Table~\ref{hbb}, 
the significance of the signal is also
shown for an integrated luminosity of 300 fb$^{-1}$. 
The $S/\sqrt{S+B}$ ratio can be as high as $5.6$ for $C_t^S = -1$, but 
however, it decreases rapidly to only $0.4$ for $C_t^S =1$.
The signal event rates are about $18- 270$ 
for $C_t^S = 1$ to $-1$ with an integrate luminosity of 300 fb$^{-1}$.

\subsubsection{$h \to \gamma\gamma$}
Diphoton decay mode of the Higgs boson is one of the two cleanest channels
of the Higgs boson, which allows a sharp reconstructed peak right at the
Higgs boson mass, and also makes the background easier to handle. The
disadvantage is that the branching ratio is small, of order $10^{-3}$, in the
SM. In this study, we employ a fixed branching ratio for $h\to\gamma\gamma$
at the SM value: $B(h \to \gamma\gamma) = 2.3\times 10^{-3}$, because there
could be extra particles running in the loop that affect the branching 
ratio. We take a conservative approach for the branching ratio 
\footnote{
Changing the value of $C_t^S$ can also affect the decay branching
ratio of $h \to \gamma \gamma$, because the decay 
proceeds via a triangular loop of the $W$ boson and the fermions
dominated by the top quark. In the SM, the contributions from the $W$ boson
and the top quark partially cancel each other. Therefore, when the sign
of the top-Yukawa is reversed, these two contributions enhance each other.
The branching ratios for $h\to \gamma\gamma$ with $C_t^S =1,0,-1$ are
$ B(h \to \gamma\gamma) = (2.3,3.7,5.4) \times 10^{-3}$, 
where we have normalized the SM value of 
$B(h \to \gamma\gamma)$ to the value given in Ref.~\cite{cern-twiki}.
}.
Furthermore, we found that it is still easier to handle the backgrounds 
with the semileptonic decay of the top quark.

\begin{figure}[tb!]
\centering
\includegraphics[width=5in]{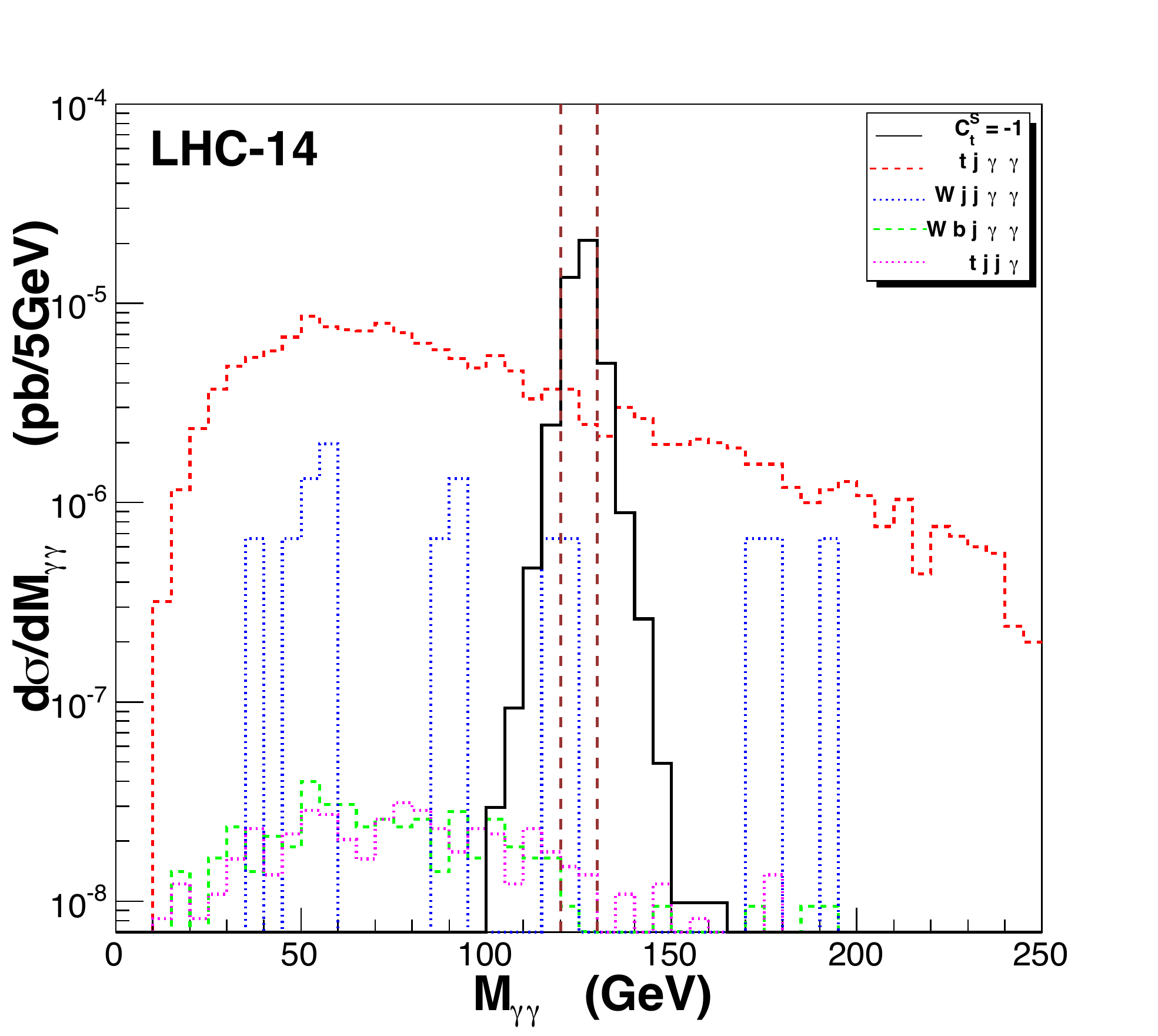}
\caption{\small \label{fig-aa}
Invariant mass $M_{\gamma\gamma}$ distributions for the signal
$pp \to t h j$ with $C_t^S = -1$ followed by the semileptonic decay
of the top quark and $h \to\gamma\gamma$, and for various backgrounds
listed above at the LHC-14. 
The shape of the peak for $C_t^S=1,0$ is the same as the
one shown with $C_t^S = -1$. 
The selection cuts up to level (3) of 
Table~\ref{hgamma} have been applied. 
The two vertical dashed lines are $m_h \pm 5$ GeV.
}
\end{figure}

The $\gamma\gamma$ decay channel has the great advantage that most 
QCD backgrounds are gone. The most relevant background comes from 
$tj \gamma\gamma$ where the photon pair is produced in the continuum. 
Inside detectors a hadronic jet sometimes can fake a photon with a
probability $O(10^{-3})$.  Therefore, $t j j\gamma$ is a background when
one of the jets fakes a photon. Other backgrounds include
$Wb j \gamma\gamma$ and $Wjj \gamma\gamma$. They are all listed in
Table~\ref{hgamma}.  
Since the spreading of the invariant-mass peak
at $m_h$ is relatively small, 
in addition to the basic cuts as in 
Eqs.~(\ref{eq:semitop-cut1}) and (\ref{eq:semitop-cut2}),
we impose the following selection cuts with the better invariant-mass window
of $\pm 5$ GeV
\begin{eqnarray}
\label{eq:semitop-cut4}
| M_{\gamma\gamma} - m_h | < 5 \; {\rm GeV}\,, \ \ 
 p_{T_\gamma} >  20\, {\rm GeV}, \ \ |\eta_\gamma|<2.5
\end{eqnarray}
to substantially reduce the background. The invariant mass distributions
for the signal and the continuum backgrounds are shown in Fig.~\ref{fig-aa}.

\begin{table}[tb!]
\caption{\label{hgamma} \small
The cut flow of 
cross sections in $10^{-3}$ fb at the LHC-14 for the signal $pp \to t h j$ 
with semileptonic decay of the top quark and $h\to \gamma\gamma$,
and various  backgrounds.  We have used a $B$-tag 
efficiency $\epsilon_b = 0.6$, mis-tag $P_{c \to b} = 0.08$
and $P_{udsg \to b} = 0.004$, and the jet-fake rate $P_{j\to \gamma} = 10^{-3}$.
We employ a fixed branching ratio for $B(h\to \gamma\gamma) = 
2.3\times 10^{-3}$.
}
\medskip
\begin{ruledtabular}
\begin{tabular}{lccccccc}
  & \multicolumn{3}{c}{Signals ($10^{-3}$ fb)} & 
              \multicolumn{4}{c}{Backgrounds ($10^{-3}$ fb) }\\
       & $C_t^S=1$ & $C_t^S=0$ & $C_t^S=-1$ & 
 $tj \gamma\gamma$ & $t j j \gamma $ & $Wb j \gamma\gamma$& $Wjj\gamma\gamma$\\
\hline 
(1) Basic cuts Eq.~(\ref{eq:semitop-cut1})
         & & & & & & &  \\
 {} and $p_{T_\gamma} > 20\, {\rm GeV}$, $|\eta_\gamma|<2.5$ & $4.45$ & $22.7$ & $80.0$ 
          & $318$ & $2.59$ & $10.5$ & $217$ \\
(2)  $2.5 < |\eta_j|<4.7$ & $2.35$ & $13.1$ & $45.2$ & $164$ & 
 $0.650$ & $1.04$ & $20.5$ \\
(3)  $M_{bl} < 200$ GeV & $2.30$ & $12.7$ & $43.6$ & $162$ & $0.609$ & 
 $0.609$ & $11.2$ \\
(4)  $|M_{\gamma\gamma} -m_h | < 5\,{\rm GeV}$ & $1.83$ & $10.2$ & 
 $34.7$ & $5.77$ & $0.027$ & $0.018$ & $0.661$ \\
\hline
$S/\sqrt{S+B}$ for 300 fb$^{-1}$ & $0.35$ & $1.4$ & $3.0$ & & & & 
\end{tabular}
\end{ruledtabular}
\end{table}

We show the cut flow of cross sections for the
signals and backgrounds  at the LHC-14 in Table~\ref{hgamma}.
The cross sections shown are calculated with $B$-tagging efficiency 
$\epsilon_b = 0.6$, mis-tag probabilities $P_{c\to b} =0.08$ and
$P_{udsg \to b} = 0.004$, and the jet-fake-photon rate of $10^{-3}$.
At the end of the set of cuts, the largest background is the
continuum of $tj\gamma\gamma$ followed by $Wjj \gamma\gamma$. 
The largest signal here is obtained with $C_t^S = -1$ at the order 
of $35\times 10^{-3}$ fb, which gives about $10$ events with an 
integrated luminosity of
300 fb$^{-1}$ while the total background has only $2$ events, although 
the signal with $C_t^S=1$ gives less than 1 event.
The significance of the signal given by $S/\sqrt{S+B}$ is also shown in
the table. Although in this $\gamma\gamma$ 
channel the ratio of $S/B$ is better than the
$b\bar b$ channel, the significance is, however, weaker because of the 
much fewer signal events.

\subsubsection{$h \to \tau^+ \tau^-$}
\begin{figure}[tb!]
\centering
\includegraphics[width=5in]{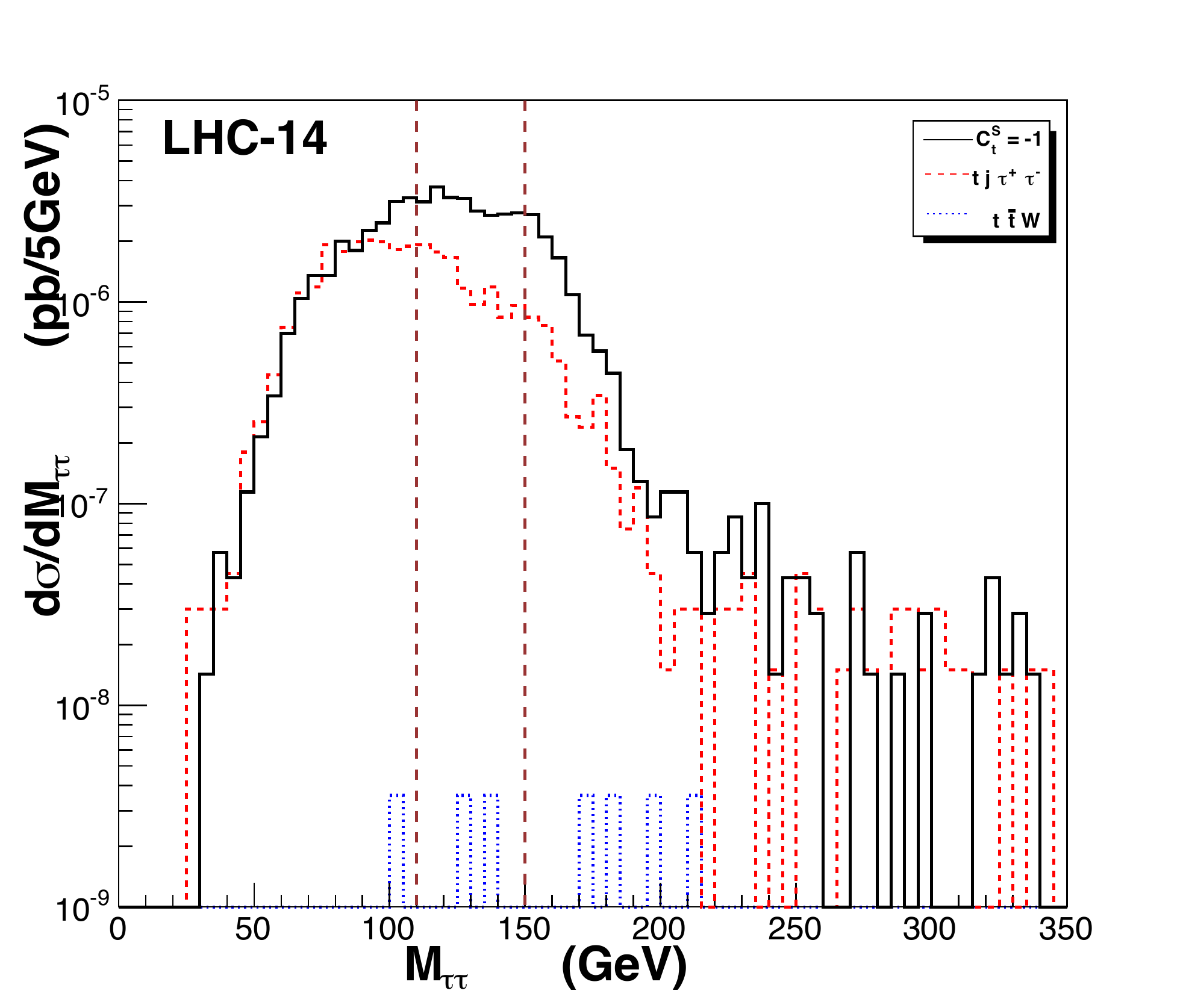}
\caption{\small \label{fig-tau}
Invariant mass $M_{\tau^+ \tau^-}$ distributions for the signal
$pp \to t h j$ with $C_t^S = -1$ followed by the semileptonic decay
of the top quark and $h \to\tau^+ \tau^-$, and for various backgrounds
listed above at the LHC-14. 
The shape of the peak for $C_t^S=1,0$ is the same as the
one shown with $C_t^S = -1$. 
The selection cuts up to level (3) of 
Table~\ref{htautau} have been applied. 
The two vertical dashed lines are at $110$ and $150$ GeV.
}
\end{figure}

The $\tau^+\tau^-$ channel has been established in the Higgs boson search
\cite{cms-tau2}.
The branching ratio for $m_h = 125-126$ GeV is about $6.2-6.3 \times 10^{-2}$
\cite{cern-twiki}. Since there are always neutrinos in tau-lepton decays, 
which means that the momentum of the parent tau lepton cannot be fully 
reconstructed. However, as the tau-lepton momentum is high enough, the
visible part of the {\it hadronic} 
tau-lepton decay can be used to determine, to
a good approximation, the parent tau-lepton momentum by a rescaling factor
(currently the tau-lepton momentum is reconstructed in the jet mode of
the tau decay, and the rescaling factor is 
$1.37$ in Delphes 3 \cite{delphes3}). 
The  reconstructed Higgs
boson peak using $\tau\tau$ channel is much broader than those
using the diphoton and 4-lepton modes: see Fig.~\ref{fig-tau}.  We 
therefore impose a loose cut in the Higgs-mass window as follows
\begin{equation}
110 \,{\rm GeV} < M_{\tau\tau} < 150 \,{\rm GeV},\;\;
 p_{T_\tau} > 25 \,{\rm GeV},\;\; |\eta_{\tau} | < 2.5 \;.
\end{equation}

The most relevant background is the continuum  $tj \tau\tau$ with
intermediate $\gamma^*$ and $Z$.
Another background is $t\bar t$ when one of the top decays hadronically
and the jets fake the tau-lepton. The $t\bar t W$ is also relevant
when $W\to \tau\nu_\tau$ and one of the top $t \to b \tau \nu_\tau$.
However, these two backgrounds turn out to be very small after cuts.

We show the cut flow of cross sections for the
signals and backgrounds at the LHC-14 in Table~\ref{htautau}.
The cross sections shown are calculated with the $B$-tagging efficiency 
$\epsilon_b = 0.6$, 
$\tau$-tagging efficiency $\epsilon_\tau=0.5$,
mis-tag probabilities $P_{c\to b} =0.08$ and
$P_{udsg \to b} = 0.004$, and the jet-fake-$\tau$ rate $P_{j\to\tau}=0.01$.
At the end of the set of cuts, the largest background is the
continuum of $tj\tau\tau$. The signal event rates are
about $0.5$ to $7$ with an integrated luminosity of 300 fb$^{-1}$ for
$C_t^S = 1$ to $-1$, and also 
the significance $S/\sqrt{S+B}$ ranges from $0.25$ to $2.3$ for $C_t^S = 1$ to
$-1$. 
The significance level is inferior to both the $b\bar b$ 
and $\gamma\gamma$ modes, mainly because of the smaller branching ratio
and the lower $\tau$ identification efficiency.

\begin{table}[tb!]
\caption{\label{htautau} \small
The cut flow of 
cross sections in fb at the LHC-14 for the signal $pp \to t h j$ 
with semileptonic decay of the top quark and $h\to \tau^+ \tau^-$,
and various backgrounds. 
We have used the $B$-tag 
efficiency $\epsilon_b = 0.6$, 
$\tau$-tagging efficiency $\epsilon_\tau=0.5$,
mis-tag $P_{c \to b} = 0.08$
and $P_{udsg \to b} = 0.004$, and the jet-fake rate $P_{j \to\tau} = 0.01$.
}
\medskip
\begin{ruledtabular}
\begin{tabular}{lcccccc}
  & \multicolumn{3}{c}{Signals (fb)} & 
              \multicolumn{2}{c}{Backgrounds (fb)}\\
       & $C_t^S=1$ & $C_t^S=0$ & $C_t^S=-1$ & 
 $tj \tau\tau$ & $t\bar t $ & $ t\bar t W$ \\
\hline 
(1) Basic cuts Eq.~(\ref{eq:semitop-cut1})
         &  & & & &   \\
 {} and $p_{T_\tau} > 25\, {\rm GeV}$, $|\eta_\tau|<2.5$ & $0.00682$ & $0.0257$ & $0.1026$  & 
$0.0701$ &  $0.420$  & $0.000672$   \\
(2)  $2.5 < |\eta_j|<4.7$ &  $0.00355$ & $0.0148$  & $0.0585$ &
 $0.0333$  & $0.0$  & $4.27\times 10^{-5}$   \\
(3)  $M_{bl} < 200$ GeV & $0.00345$  & $0.0141$  & $0.0555$   & 
$0.0319$  &  $0.0$   & $4.27\times10^{-5}$  \\
(4)  $110 < M_{\tau\tau}< 150\,{\rm GeV}$ & $0.00158$ & $0.00616$ & 
  $0.0244$ & $0.0105$ & $0.0$  & $1.904 \times 10^{-5}$ \\
\hline
$S/\sqrt{S+B}$ for 300 fb$^{-1}$ & $0.25$ & $0.83$  & $2.3$ & & &
\end{tabular}
\end{ruledtabular}
\end{table}

\subsection{Hadronic top decay}
In this subsection, we consider 
associated Higgs production with a single
top quark and a forward jet, followed by the hadronic decay of the top
quark:
\begin{equation}
pp \longrightarrow thj \longrightarrow \, (bj_1j_2)\,h\,j \;,
\end{equation}
where we label $j_1, j_2$ for the 2 jets from the top-quark decay.
We first apply the same basic cuts as in Eq.~(\ref{eq:semitop-cut1}) 
on identifying the $b$ quark and the 2 jets from the top decay, as
well as the forward jet, which has the further requirement 
\[ 
 2.5 < | \eta_j | < 4.7 \;.
\]
We also impose the requirement on the $b$ quark and the two jets that 
originate from the top decay by
\[
 M_{b j_1 j_2} < 300 \; {\rm GeV} \;.
\]

\subsubsection{$h\to Z Z^* \to 4 \ell$}
\begin{figure}[tb!]
\centering
\includegraphics[width=5in]{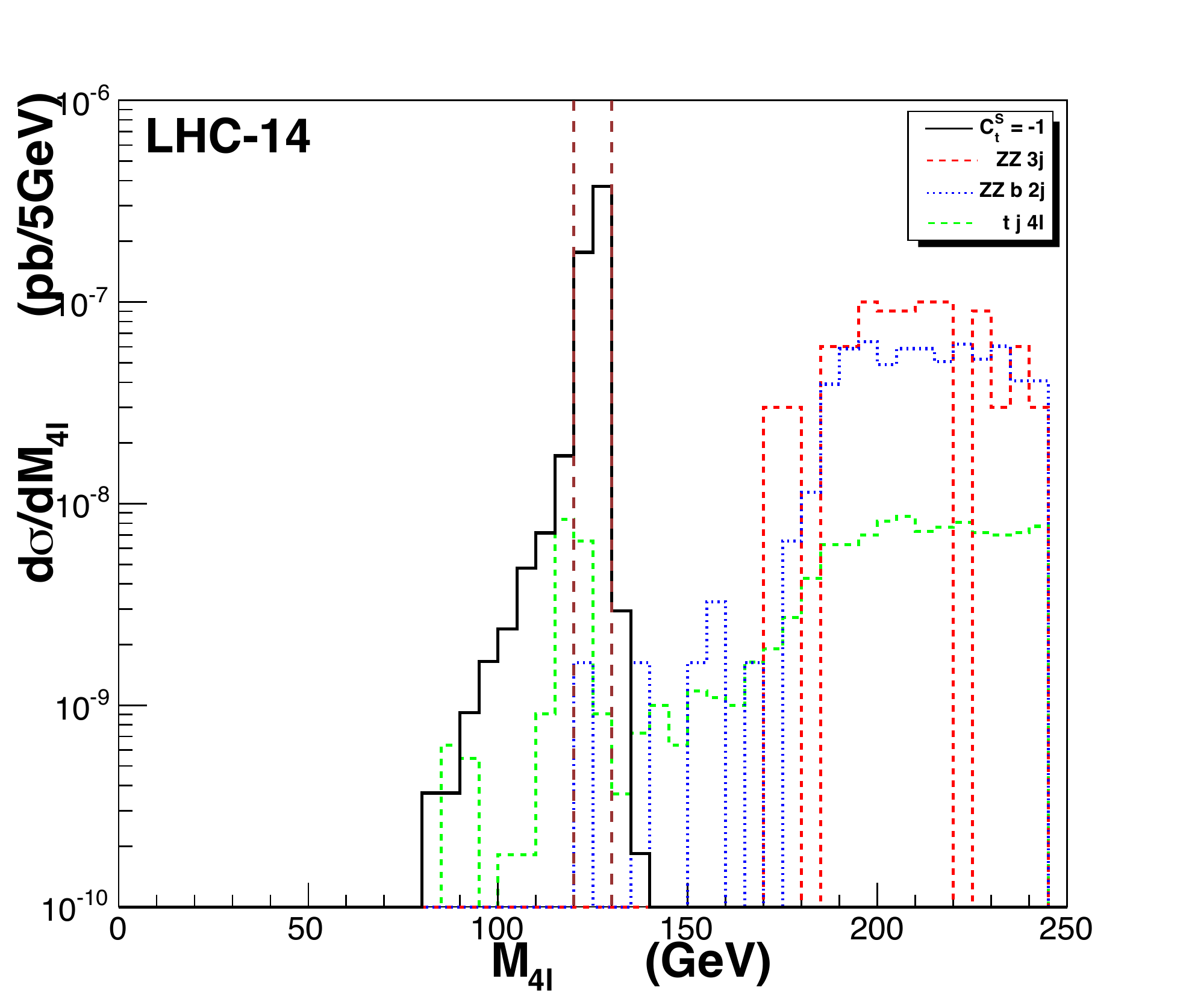}
\caption{\small \label{fig-zz}
Invariant mass $M_{4\ell}$ distributions for the signal
$pp \to t h j$ with $C_t^S = -1$ followed by the hadronic decay
of the top quark and $h \to ZZ^* \to 4 \ell $, and for various backgrounds
listed above at the LHC-14. 
The shape of the peak for $C_t^S=1,0$ is the same as the
one shown with $C_t^S = -1$. 
The selection cuts up to level (3) of 
Table~\ref{hzz} have been applied. 
The two vertical dashed lines are $m_h \pm 5$ GeV
}
\end{figure}

The Higgs boson decaying into $ZZ^* \to 4 \ell$ is one of the cleanest
channels for discovery and reconstruction. Since the branching ratio into
$4\ell$ is very small and the $p_T$ of the electrons or muons
is only of order $20$ GeV, and so we apply mild cuts for the charged 
leptons
\begin{equation}
p_{T_{\ell}} > 5 \,{\rm GeV}, \qquad
|\eta_\ell| < 2.5 \;.
\end{equation}
We further apply the Higgs-mass window cut on the
invariant mass formed by the 4 charged leptons
\begin{equation}
| M_{4\ell}  - m_h | < 5\,{\rm GeV} \;.
\end{equation}
The invariant mass distributions for the signal and various backgrounds
are shown in Fig.~\ref{fig-zz}.

We show the cut flow of cross sections for the 
signals and backgrounds at the LHC-14 in Table~\ref{hzz}.
The cross sections shown are calculated with $B$-tagging efficiency 
$\epsilon_b = 0.6$, mis-tag probabilities $P_{c\to b} =0.08$ and
$P_{udsg \to b} = 0.004$.
At the end of the set of cuts, the largest background comes from $tj4 \ell$,
where the $\ell\ell\ell\ell$ 
comes from the $\gamma^*$ and $Z^*$ exchanges, but it
is rendered extremely small. However, the signal event rates are also very
tiny, substantially smaller than 1 event for an integrated luminosity of 
300 fb$^{-1}$. Nevertheless, if integrated luminosity can increase to 
3000 fb$^{-1}$ we can have $2-3$ events for $C_t^S=-1$. The event rate is
small simply because of the tiny branching ratio of $h\to ZZ^* \to 4 \ell$.
One perhaps can perform the calculation using the $ZZ^* \to \ell^+ \ell^- j j$
mode, but we shall delay this channel in future works.

\begin{table}[bth!]
\caption{\label{hzz} \small
The cut flow of 
cross sections in $10^{-3}$ fb at the LHC-14 for the signal $pp \to t h j$ 
with hadronic decay of the top quark and $h\to ZZ^* \to 4 \ell$,
and various backgrounds. 
 We have used a $B$-tag 
efficiency $\epsilon_b = 0.6$, mis-tag $P_{c \to b} = 0.08$
and $P_{udsg \to b} = 0.004$.
}
\medskip
\begin{ruledtabular}
\begin{tabular}{lcccccc}
  & \multicolumn{3}{c}{Signals ($10^{-3}$ fb)} & 
              \multicolumn{2}{c}{Backgrounds ($10^{-3}$ fb)}\\
       & $C_t^S=1$ & $C_t^S=0$ & $C_t^S=-1$ & 
 $tj 4 \ell $ & $ZZ 3j $ & $ZZb 2j$ \\
\hline 
(1) Basic cuts Eq.~(\ref{eq:semitop-cut1}) & & & & &  & \\
{} and $p_{T_{j_{1,2}}} > 25$ GeV, $|\eta_{j_{1,2}}|<2.5$
& & & & &  & \\
{} but with $p_{T_{\ell}} > 5\, {\rm GeV}$ & 
$0.136$ & $0.531$ & $1.77$  & $0.955$ &  $20.1$  & $10.0$   \\
(2)  $2.5 < |\eta_j|<4.7$ &  $0.091$ & $0.366$  & $1.18$ &
 $0.539$  & $8.01$  & $5.01$   \\
(3)  $M_{b j_1 j_2} < 300$ GeV & $0.081$  & $0.324$  & $1.02$   & 
$0.438$  &  $3.79$   & $1.97$  \\
(4)  $| M_{4\ell}  - m_h | < 5\,{\rm GeV}$ & $0.072$ & $0.289$ & 
  $0.901$ & $8.65 \times 10^{-3}$ & $0.0$  & $0.0$ \\
\hline
$S/\sqrt{S+B}$ for 300 fb$^{-1}$ & $0.14$ & $0.29$  & $0.52$ & & &
\end{tabular}
\end{ruledtabular}
\end{table}

\subsection{Distinction among $C_t^S = 1,0,-1$}
\begin{figure}[tbh!]
\centering
\includegraphics[width=3.2in]{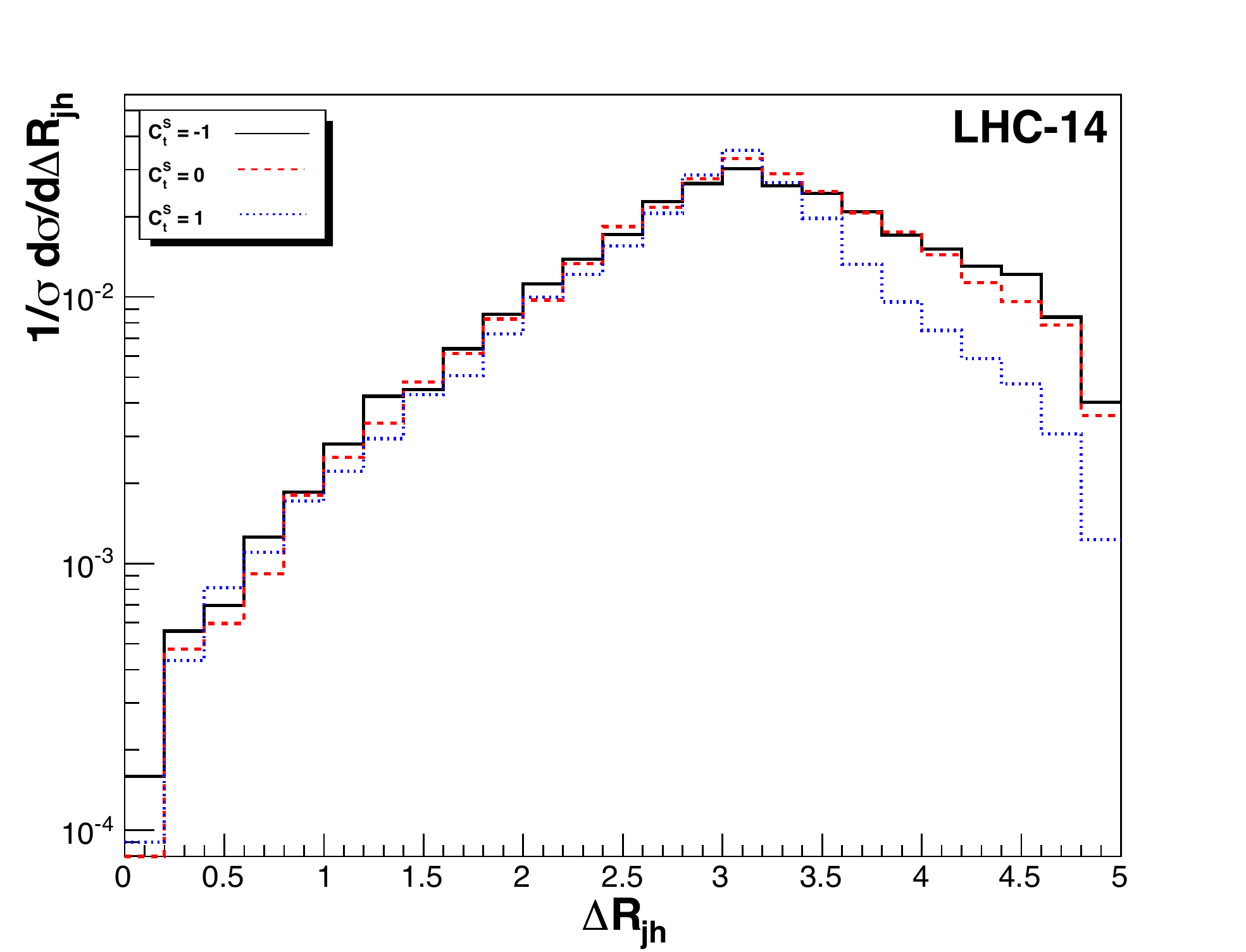}
\includegraphics[width=3.2in]{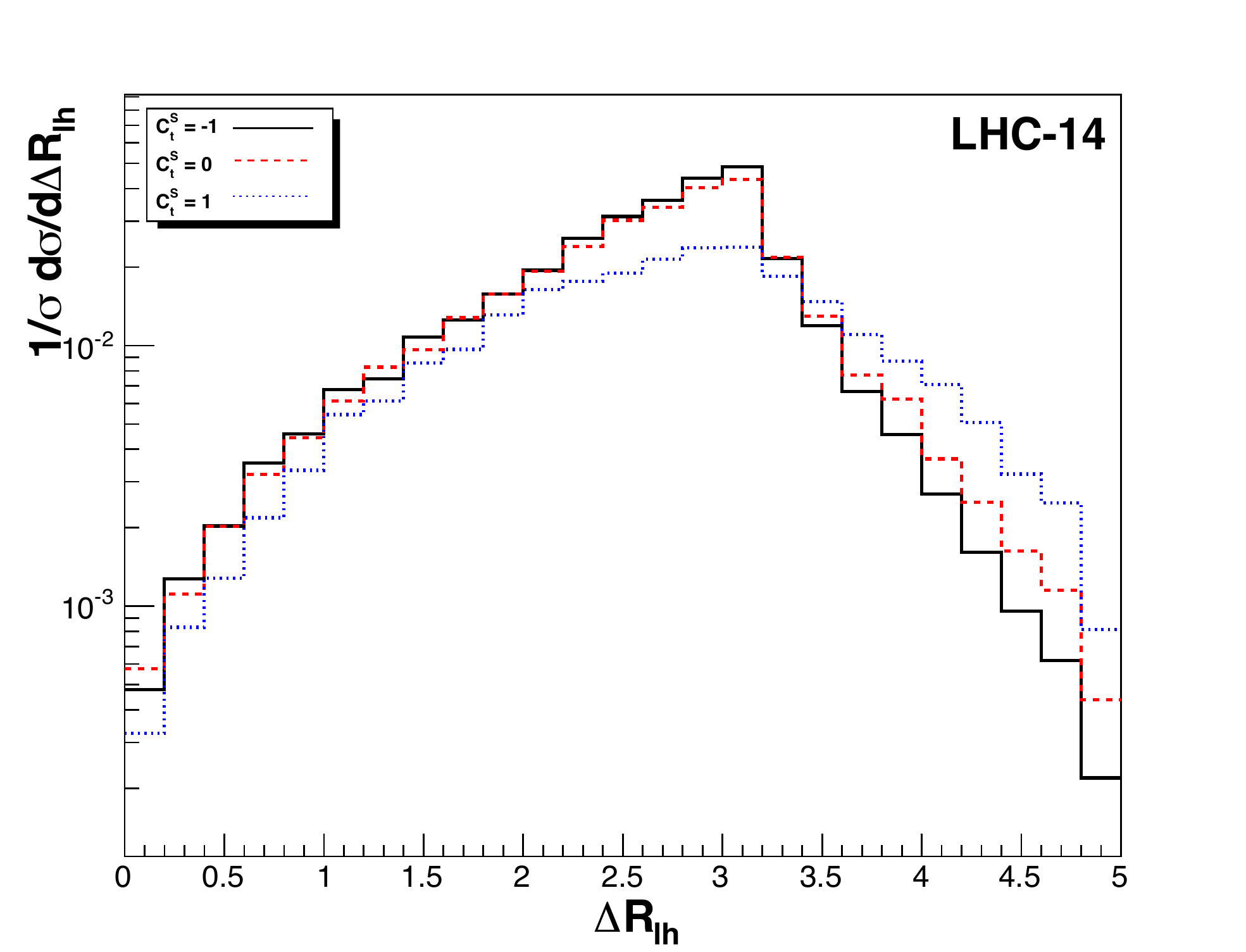}
\includegraphics[width=3.2in]{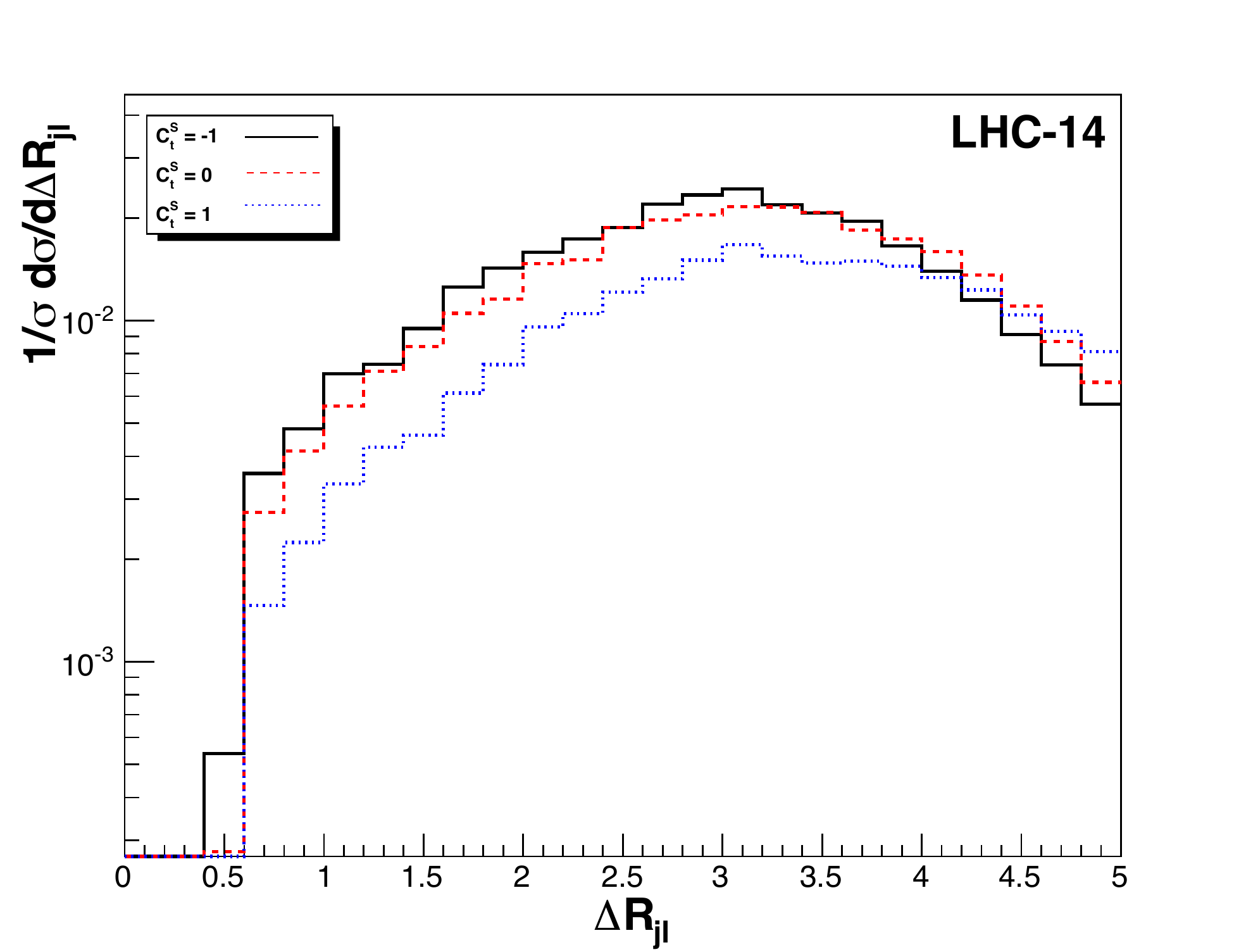}
\caption{\small \label{fig-dR}
Normalized $\Delta R$ distributions for various pairs of particles $(\ell,j,h)$,
where the momentum of $h$ is reconstructed by the photon pair, for
the signal process  $pp \to t h j$ with $C_t^S = -1,0,1$ followed by the 
semileptonic decay of the top quark and $h \to \gamma\gamma $
at the LHC-14. 
Behavior of $b$ and $\ell$ is about the same, as they are coming
from the same top quark decay. We need only one of them: $\ell$.
}
\end{figure}

In addition to the differences in cross section, 
we further found that the spatial separations $\Delta R$ among the forward
jet, the $b$ quark, the charged lepton, and the reconstructed Higgs boson
show interesting differences among 
$C_t^S = 1,0,-1$, as shown in Fig.~\ref{fig-dR}.
Without loss of generality we use the $h\to\gamma\gamma$ decay mode for
this study, because the 4-momentum of the Higgs boson can be reconstructed
cleanly by summing the 4-momenta of the two photons.
Since the behavior of the charged lepton and the $b$ quark coming
from the top quark decay is similar, we choose only the charged lepton 
to show the $\Delta R$ distributions.
We found that the spatial separation $\Delta R$ between the forward jet
and the Higgs boson becomes wider when $C_t^S$ deviates from the SM value $1$,
while that between the charged lepton and the Higgs boson  and that
between the forward jet and the charged lepton become narrower as $C_t^S$
deviates from $1$. 
Similar patterns were observed in Ref.~\cite{Englert:2014pja}.
It was also shown in Ref.~\cite{ellis} that the variations in 
scalar and pseudoscalar components of the top-Yukawa coupling can also induce
interesting angular correlations.

\section{Discussion}
\begin{figure}[th!]
\centering
\includegraphics[width=5in]{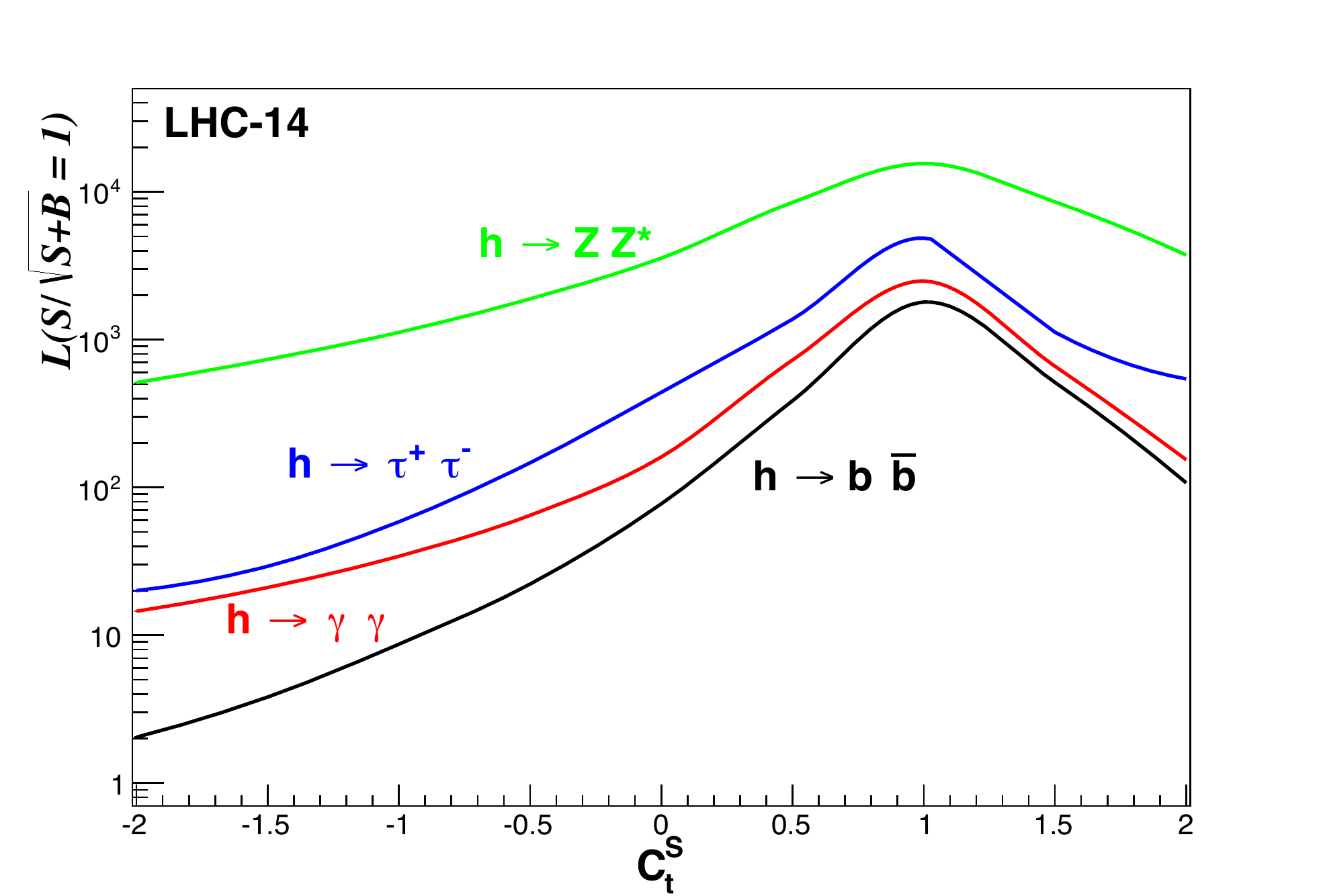}
\caption{\small \label{lumin}
Required luminosities at the LHC-14 for the process $pp \to thj$ 
in various decay channels of the Higgs boson to
achieve $S/\sqrt{S+B} = 1$.  We show the channels
$h\to b\bar b$, $\gamma\gamma$, $\tau^+\tau^-$, and $ZZ^* \to 4\ell$.
}
\end{figure}

We have studied associated Higgs production with a single top quark
in the dominant process $pp \to thj$, followed by the semileptonic
decay of the top quark and $h \to b\bar b,\, \gamma\gamma,\,
\tau^+ \tau^-$ and by the hadronic 
decay of the top quark and $h \to ZZ^* \to 4\ell$.  
So far, we have
found that the $h \to b \bar b$ channel offers the best 
chance in terms of 
significance $S/\sqrt{S+B}$ for observing the signal with various $C_t^S$.
When $C_t^S =1$ (SM) the significance level is very low at $0.4$, but it quickly
rises to a large enough value $5.6$ when $C_t^S= -1$. The signal event
rates are from 
$18$ to $270$ with an integrated luminosity of 300 fb$^{-1}$.
In Fig.~\ref{lumin}, we show the required luminosities to achieve
a significance level of $S/\sqrt{S+B} = 1$ for various channels considered
in this work versus $C_t^S = -2$ to  $2$.  
Note that $S/\sqrt{S+B} > 1$ implies
\[
  S > 1 + \frac{B}{S} > 1
\]
i.e., the event rate $S>1$ is guaranteed. 
The best channel is the
$h\to b \bar b$. The second and the third are $h\to \gamma\gamma$ and 
$h\to \tau^+ \tau^-$, respectively. The last one is
the $h\to ZZ^* \to 4\ell$. 
Note that the  $ZZ^*$ channel requires a very large
luminosity in order to achieve $S/\sqrt{S+B} > 1$ simply because
of its very small signal cross sections.
Similarly, 
because of the small signal cross sections in $\gamma\gamma$ channel,
the $h\to\gamma\gamma$ channel requires larger luminosities
than the $b\bar b$ channel in order to achieve $S/\sqrt{S+B} > 1$,
although the $S/B$ ratio is much better in the
$\gamma\gamma$ than in the $b\bar b$ channel.

Before we close we offer the following comments.
\begin{enumerate}
\item In the current framework, the bottom-Yukawa has very small effects
on the cross section, because the bottom-Yukawa coupling $C_b^S$
is approximately constrained to be within the range $-2$ to $+2$ 
\cite{higgcision}.

\item The higher order process, $pp \to th bj$, contains one more $b$ quark
in the final state.  Potentially, it can increase the signal sensitivity
based on a parton-level study \cite{farina}. However, we have shown in this 
work that with full detector simulations the Higgs-mass window cut is not
as effective as that in parton-level. Furthermore, there are further
combinatorics problems as we have to identify the $b$ quark from top decay
and the two $b$ quarks from the Higgs boson decay.

\item The  $h\to b\bar b$ decay mode turns out to be the best 
in terms of significance to probe
associated Higgs production with a single top quark, 
because of the larger event rates, although the signal-to-background
ratio is less than 1.

\item The best signal-to-background ratio $S/B$ is achieved in the
$h\to Z Z^* \to 4\ell$ channel, followed by the $\gamma\gamma$,
$\tau^+ \tau^-$, and $b\bar b$ channels. Nevertheless, the event rates
in $4\ell$ and $\gamma\gamma$ are too low for detection.

\item Better $\tau$-lepton identification is needed in order to raise 
the efficiency in identifying the $h\to \tau\tau$ decay. Perhaps, one
can look into the substructure in the fast-moving $\tau$ jet. 
If one can achieve better efficiencies, it will enhance the probe of
the single top associated Higgs production.

\item The $h \to ZZ^* \to 4\ell$ has a very small branching ratio, and so
the detection of single top associated Higgs production requires
an extremely high luminosity. One should pursue the $ZZ^* \to \ell^+ \ell^-
jj$ mode, which has about one order of magnitude larger in event rates, but
different backgrounds. 

\item We do not attempt the $h\to WW^*$ mode in this work, simply 
because the Higgs boson peak cannot be reconstructed in this mode,
unless we go for the $4j$ mode.

\item If the top-Yukawa is close to the SM value, the best chance to
observe associated Higgs production with a single top quark is via
the $h\to b \bar b$ channel. However, it requires an integrated 
luminosity more than 1000 fb$^{-1}$.

\end{enumerate}
In summary, we have studied the effects of varying the top-Yukawa coupling 
in the Higgs associated production with a single top quark,
with full detector simulations,
We found that the $h\to b\bar b$ mode
with the semileptonic decay of the top quark has the highest potential
in observing the process and the effects of top-Yukawa coupling,
especially the sign of the top-Yukawa coupling can be determined.

\section*{Acknowledgment}  
This work was supported by the National Science
Council of Taiwan under Grants No. 102-2112-M-007-015-MY3,
and by
the National Research Foundation of Korea (NRF) grant
(No. 2013R1A2A2A01015406).
This study was also
financially supported by Chonnam National University, 2012.
J.S.L thanks National Center for Theoretical Sciences (Hsinchu, Taiwan) for the 
great hospitality
extended to him while this work was being performed.

\section*{Appendix}

\def\theequation{A.\arabic{equation}}
\begin{appendix}
\setcounter{equation}{0}
In this appendix, we present the amplitude of the process $qb \to t h q' $
in the effective $W$ approximation assuming $h$ is a spin-zero CP-mixed state. 
In this process, the dominant contribution comes from the region where
the $W$ boson emitted from the incoming quark $q$ is close to onshell and
one can approximately represent the process by
the $W$ boson scattering with the incoming $b$ quark 
to give $h$ and $t$ in the final state:
\begin{equation}
W (p_W) \ b(p_b) \ \to \ h(p_h) \ t (p_t) \,.
\end{equation}
The process $W b \to h t$ receives contributions from
$(a)$ a $t$-channel diagram with the $W$ exchange and 
$(b)$ a $s$-channel diagram with the $t$ exchange. 
The vertex factor for $hWW$ is given in Eq.~(\ref{hvv}) 
and that for $ht\bar t$ in Eq.~(\ref{hff}) with the identification
of $g_{hWW} = C_v$ and $g^{S,P}_{htt} = C^{S,P}_t$,
see Eq.~(\ref{eq:notation}).
Then, the amplitude of each diagram reads
\begin{eqnarray}
{\cal M}_{(a)}&=&\frac{g^2m_t C_v}{\sqrt{2}m_W (t-m_W^2)}\,
\left[
(p_b-p_t)\cdot\epsilon(p_W)\,\bar{u}(p_t) P_L u(p_b) \ + \
\frac{m_W^2}{m_t}\,
\bar{u}(p_t) \slash\epsilon(p_W) P_L u(p_b) \right]\,,  \\[3mm]
{\cal M}_{(b)}&=&\frac{g^2m_t }{\sqrt{2}m_W (s-m_t^2)}\,
\left[
m_t C_t^S \,\bar{u}(p_t) \slash\epsilon(p_W) P_L u(p_b) \ + \
\left(\frac{C_t^S-iC_t^P}{2}\right)\,
\bar{u}(p_t) {\not\!p}_h  \;
\slash\epsilon(p_W) P_L u(p_b) \right]\,, \nonumber
\end{eqnarray}
where $s=(p_b+p_W)^2=(p_t+p_h)^2$,
$t=(p_b-p_t)^2=(p_h-p_W)^2$, and
$u=(p_b-p_h)^2=(p_t-p_W)^2$ and $\epsilon^\mu(p_W)$ denotes 
the polarization vector of $W$ boson.
In the high-energy limit of $s,|t|,|u| \gg m_W^2,m_h^2,m_t^2$, we find that
\begin{equation}
{\cal M}={\cal M}_{(a)}+{\cal M}_{(b)} \approx -\,
\frac{g^2m_t}{2\sqrt{2}m_W^2}\,\left[(C_v-C_t^S)+iC_t^P\right]\,
\bar{u}(p_t) P_L u(p_b)
\end{equation}
taking the longitudinally polarized $W$ or $\epsilon^\mu(p_W)\approx p_W^\mu/m_W$
and using $(p_b-p_t)\cdot\epsilon(p_W) \approx -t/(2m_W)$ and
$\bar{u}(p_t)\, {\not\!p}_h \; 
\slash\epsilon(p_W) P_L u(p_b)\approx (s/m_W)\,\bar{u}(p_t) P_L u(p_b)$.
We observe our results are consistent with those given in 
Ref.~\cite{tim}.
We note that, in the high-energy limit,
\begin{equation}
\overline{\left|{\cal M}\right|^2}\propto
\left[(C_v-C_t^S)^2+(C_t^P)^2\right]\,(-t )
\end{equation}
and therefore the absence of this
unitarity-breaking term requires $C_v=C_t^S$ and $C_t^P=0$.

\end{appendix}


\end{document}